\documentclass[acmsmall, review]{acmart}
\usepackage{libertine}
\usepackage{microtype}
\usepackage{booktabs} 
\usepackage{comment} 
\usepackage{paralist}
\setdefaultenum{1)}{}{}{}
\usepackage{amsmath}

\usepackage{graphicx}
\usepackage{subcaption}
\usepackage{adjustbox}
\usepackage{listings}

\usepackage{colortbl}
\usepackage{multicol}
\usepackage{multirow}
\usepackage{bm}

\usepackage{changebar}

\usepackage[ruled,vlined]{algorithm2e}

\DontPrintSemicolon
\SetKwComment{tcp}{$\triangleright$ }{}
\SetVlineSkip{0cm}
\SetKwInOut{Param}{Param}
\SetKwInOut{Input}{Input}

\usepackage{tikz}
\usetikzlibrary{shapes,shapes.geometric,arrows.meta,fit,calc,positioning,chains}

\newcommand{\TODO}[1]{\textcolor{red}{\large\textbf{#1}}}

\definecolor{lightgray}{gray}{0.9}
\renewenvironment{thebibliography}[1]{%
   \begin{oldthebibliography}{#1}%
     \setlength{\itemsep}{-.3ex}%
}%
{%
   \end{oldthebibliography}%
}

\usepackage[compact]{titlesec}
\titlespacing{\section}{2pt}{*0}{*0}
\titlespacing{\subsection}{1pt}{*0}{*0}

\acmJournal{TOPC}

\begin{document}
\title{Programming Strategies for Irregular Algorithms on the Emu Chick}
\author{Eric Hein}
\affiliation{~Emu Technology}
\author{Srinivas Eswar, Abdurrahman Ya\c{s}ar}
\affiliation{~Georgia Institute of Technology}
\author{Jiajia Li}
\affiliation{~Pacific Northwest National Laboratory}
\author{Jeffrey S. Young}
\author{Thomas M. Conte}
\author{\"{U}mit V. \c{C}ataly\"{u}rek}
\author{Rich Vuduc}
\author{Jason Riedy}
\orcid{0000-0002-4345-4200}
\affiliation{~Georgia Institute of Technology}
\author{Bora U\c{c}ar}
\affiliation{~CNRS and LIP, \'{E}cole Normale Sup\'{e}rieure de Lyon}





\maketitle
\thispagestyle{plain}
\pagestyle{plain}


\section*{Abstract}\label{sec:abstract}
The Emu Chick prototype implements migratory memory-side processing in a novel hardware system. Rather than transferring large amounts of data across the system interconnect, the Emu Chick moves lightweight thread contexts to near-memory cores before the beginning of each remote memory read. Previous work has characterized the performance of the Chick prototype in terms of memory bandwidth and programming differences from more typical, non-migratory platforms, but there has not yet been an analysis of algorithms on this system.

This work evaluates irregular algorithms that could benefit from the lightweight, memory-side processing of the Chick and demonstrates techniques and optimization strategies for achieving performance in sparse matrix-vector multiply operation (SpMV), breadth-first search (BFS), and graph alignment across up to eight distributed nodes encompassing 64 nodelets in the Chick system.
We also define and justify relative metrics to compare prototype FPGA-based hardware with established ASIC architectures.
The Chick currently supports up to 68x scaling for graph alignment, 
80 MTEPS for BFS on balanced graphs, and 50\% of measured STREAM bandwidth for SpMV.



\section{Introduction}
\label{sec:intro}

\begin{figure}
\centering
\includegraphics[width=0.8\linewidth]{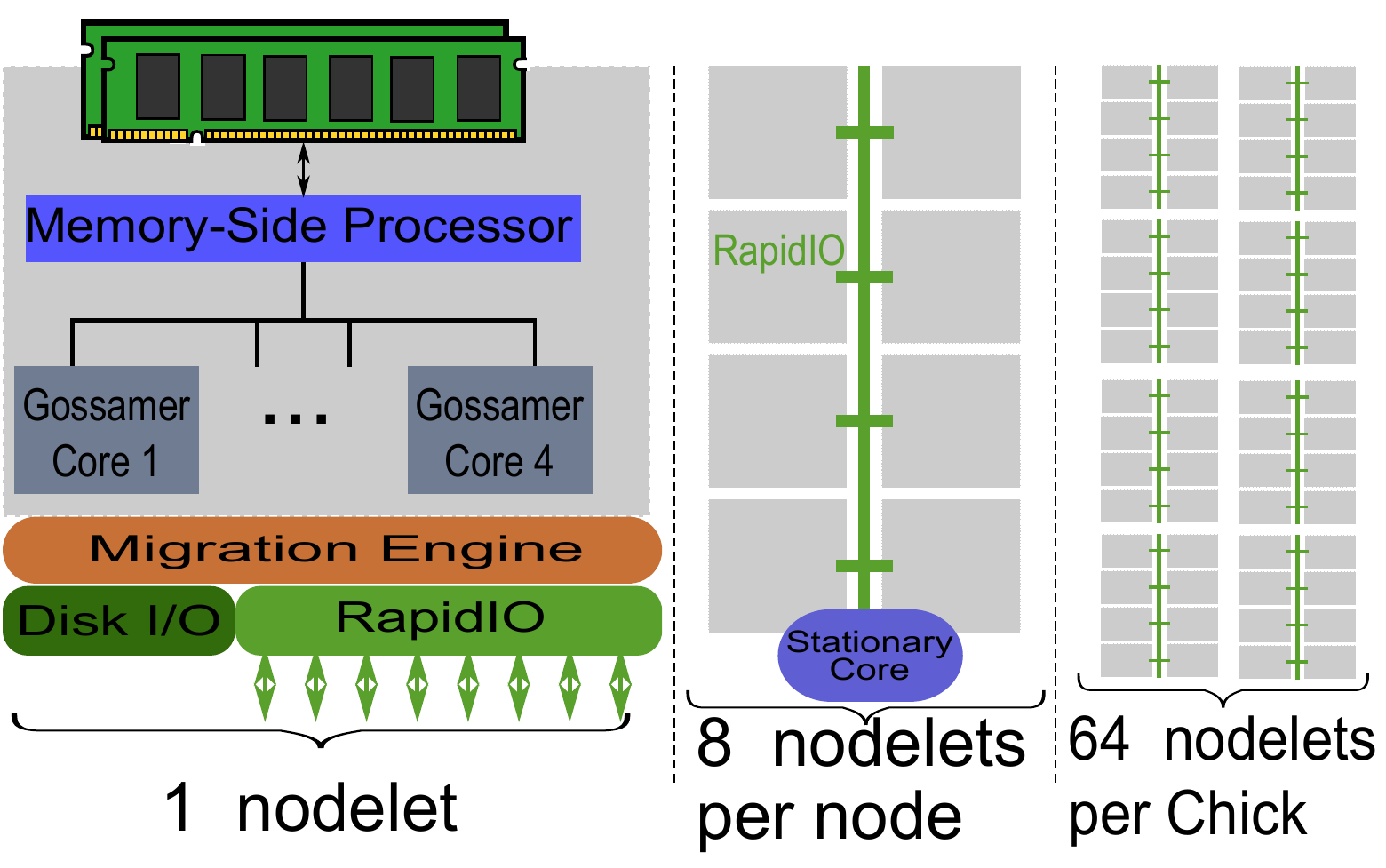}
\caption{{\small{Emu architecture: The system consists of \emph{stationary} processors for running the operating system and up to four \emph{Gossamer} processors per nodelet tightly coupled to memory.  The cache-less Gossamer processing cores are multi-threaded to both source sufficient memory references and also provide sufficient work with many outstanding references.  The coupled memory's narrow interface ensures high utilization for accesses smaller than typical cache lines.}}}\label{fig:emu-arch}
\end{figure}

Analyzing data stored in irregular data structures such as graphs and sparse matrices
is challenging for traditional architectures due to limited data locality in associated algorithms and performance costs related to data movement. The Emu architecture~\cite{dysart2016emu} is designed specifically to address these data movement costs in a power-efficient hardware environment by using a cache-less system built around ``nodelets'' (see Figure~\ref{fig:emu-arch}) that execute lightweight threads. These threads migrate on remote data reads rather than pulling data through a traditional cache hierarchy. The key differentiators for the Emu architecture are the use of cache-less processing cores, a high-radix network connecting distributed memory, and PGAS-based data placement and accesses. In short, the Emu architecture is designed to scale applications with poor data locality to supercomputing scale by more effectively utilizing available memory bandwidth and by dedicating limited power resources to networks and data movement rather than caches.  

Previous work has investigated the initial Emu architecture design \cite{dysart2016emu}, algorithmic designs for merge and radix sorts on the Emu hardware \cite{minutoli2015implementing}, and baseline performance characteristics of the Emu Chick hardware \cite{hein2018emuchar,belviranli:2018:emuhpec}. This investigation is focused on determining how irregular algorithms perform on the prototype Chick hardware and how we implement specific algorithms so that they can scale to a rack-scale Emu and beyond.

This study's specific demonstrations include:
\begin{compactitem}
\item The first characterization of the Emu Chick hardware using irregular algorithms including sparse matrix vector multiply (SpMV), graph analytics (BFS), and graph alignment. We also discuss programming strategies for the Emu such as \textit{replication} (SpMV), \textit{remote writes to reduce migration} (BFS), and \textit{data layout to reduce workload imbalance} (graph alignment) that can be used to increase parallel performance on the Emu. 
\item Multi-node Emu results for BFS scaling up to 80 MTEPS and 1.28 GB/s on a balanced graph as well as an initial comparison of Emu-optimized code versus a naive Cilk implementation on x86.
\item Multi-node results for SpMV scaling up to 50\% of measured peak bandwidth on the the Emu. 
\item Graph alignment results showing a 68x speedup when scaling from 1 to 256 threads on 8 nodelets with optimized data layout and comparison strategies. 
\end{compactitem}


\section{The Emu Architecture}
\label{sec:background}


The Emu architecture focuses on improved random-access bandwidth scalability by migrating lightweight \emph{Gossamer} threads, or ``threadlets'', to data and emphasizing fine-grained memory access.
A general Emu system consists of the following processing elements, as illustrated in Figure~\ref{fig:emu-arch}:
\begin{compactitem}
\item A common \emph{stationary} processor runs the OS (Linux) and manages storage and network devices.
\item \emph{Nodelets} combine narrowly banked memory with highly multi-threaded, cache-less \emph{Gossamer} cores to provide a memory-centric environment for migrating threads.
\end{compactitem}
These elements are combined into nodes that are connected by a RapidIO fabric. The current generation of Emu systems include one stationary processor for each of the eight nodelets contained within a node. System-level storage is provided by SSDs. We talk more specifically about some of the prototype limitations of our Emu Chick in Section~\ref{sec:exp}. More detailed descriptions of the Emu architecture are available~\cite{dysart2016emu}, but this is a point in time description of the current implementation and its trade-offs.

For programmers, the Gossamer cores are transparent accelerators. The compiler infrastructure compiles the parallelized code for the Gossamer ISA, and the runtime infrastructure launches threads on the nodelets. Currently, one programs the Emu platform using Cilk~\cite{leiserson1997programming}, providing a path to running on the Emu for OpenMP programs whose translations to Cilk are straightforward.
The current compiler supports the expression of task or fork-join parallelism through Cilk's \texttt{cilk\_spawn} and \texttt{cilk\_sync} constructs, with a future Cilk Plus (Cilk+) software release in progress that would include \texttt{cilk\_for} (the nearly direct analogue of OpenMP's \texttt{parallel for}) as well as Cilk+ reducer objects. Many existing C and C++ OpenMP codes can translate almost directly to Cilk+.

A launched Gossamer thread only performs local reads. Any remote read triggers a migration, which will transfer the context of the reading thread to a processor local to the memory channel containing the data. Experience on high-latency thread migration systems like Charm++ identifies migration overhead as a critical factor even in highly regular scientific codes~\cite{7013040}. The Emu system minimizes thread migration overhead by limiting the size of a thread context, implementing the transfer efficiently in hardware, and integrating migration throughout the architecture. In particular, a Gossamer thread consists of 16 general-purpose registers, a program counter, a stack counter, and status information, for a total size of less than 200 bytes. The compiled executable is replicated across the cores to ensure that instruction access always is local. Limiting thread context size also reduces the cost of spawning new threads for dynamic data analysis workloads. Operating system requests are forwarded to the stationary control processors through the service queue.

The highly multi-threaded Gossamer cores read only local memory and do not have caches, avoiding
cache coherency traffic.
Additionally, ``memory-side processors'' provide atomic read or write operations that can be used to access small amounts of data without triggering unnecessary thread migrations. A node's memory size is relatively large with standard DDR4 chips (64\,GiB) but with multiple, Narrow-Channel DRAM (NCDRAM)  memory channels (8 channels with 8 bit interfaces to the host using FIFO ordering). Each DIMM has a page size of 512B and a row size of 1024. The smaller bus means that each channel of NCDRAM has only 2GB/s of bandwidth, but the system makes up for this by having many more independent channels. Because of this, it can sustain more simultaneous fine-grained accesses than a traditional system with fewer channels and the same peak memory bandwidth.


\section{Algorithms}
\label{sec:algorithms}
We investigate programming strategies for three algorithms: 1) the standard (CSR) sparse matrix vector multiplication operation, 2) Graph500's breadth-first search (BFS) benchmark, and 3) graph alignment, computing a potential partial mapping of the vertices of two graphs.
These three algorithms cover a variety of sparse, irregular computations: the ubiquitous sparse matrix vector multiplication, filtered sparse matrix sparse vector multiplication (in BFS), and a variant of the sparse matrix - sparse matrix multiplication (in computing the similarities of vertices). In the following subsections we discuss how we implement these algorithms on the Emu platform.

\subsection{Sparse Matrix Vector Multiply (SpMV):}



\begin{figure}
  \centering
  \includegraphics[width=0.4\linewidth]{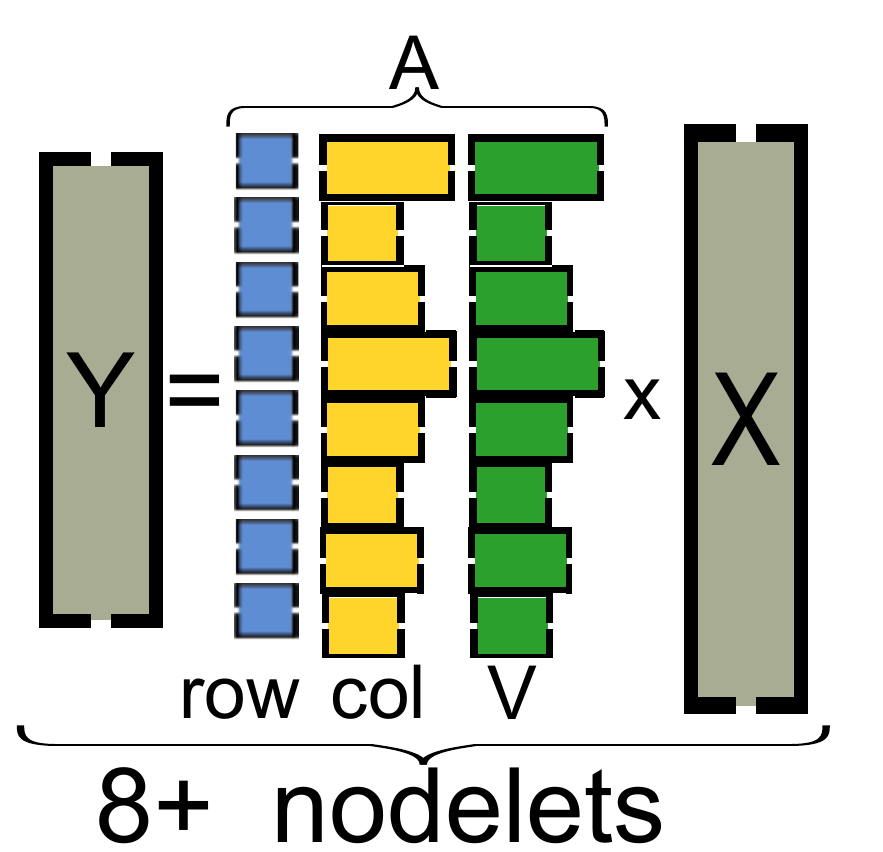}
  \caption{{\small{Emu-specific layout for CSR SpMV.}}}
  \label{fig:csr_spmv_layout}
\end{figure}

The matrix $A$ is stored as a distributed CSR structure consisting of 3 arrays - row offsets, column indices, and values. The row offset array is striped across all nodelets and encodes the length of each row. Every row's non-zero entries and column indices are allocated together and are present in the same nodelet giving rise to the jagged arrays \texttt{col} and \texttt{V} shown in Figure \ref{fig:csr_spmv_layout}. \texttt{X} is replicated across each nodelet and the output \texttt{Y} is striped across all nodelets.

In the 2D allocation case, no thread migrations occur when accessing elements in the same row.
A 1D striped layout incurs a migration for every element within a row to fetch the vector entry.
Synthetic Laplacian matrix inputs are created corresponding to a $d$-dimensional $k$-point stencil on a grid of length $n$ in each dimension. For the tested synthetic matrices, $d=2$ and $k=5$, resulting in a $n^2 \times n^2$ Laplacian with five diagonals.  The tested real world matrices are listed in Table~\ref{table:spmv-real-multinode}.

\subsection{Graph Analytics (Breadth First Search for Graph500)}
  \label{ssec:bfs_emu_alg}

Our in-memory graph layout is inspired by STINGER \cite{stinger-hpec12} so that computation can adapt to a changing environment\cite{mlg2018_23}. Each vertex contains a pointer to a linked-list of edge blocks, each of which stores a fixed number of adjacent vertex IDs and a pointer to the next edge block. We use a striped array of pointers to distribute the vertex array across all nodelets in the system, such that vertex 0 is on nodelet 0, vertex 1 is on nodelet 1, and so on.
We use STINGER rather than CSR to enable future work with streaming data and incremental algorithms\cite{dynograph}, one of the primary targets of the Emu architecture.
Note that breadth-first search is nearly equivalent to computing a filtered sparse matrix times sparse vector product \cite{oai:arXiv.org:1606.05790}.

To avoid the overhead of generic run-time memory allocation via \texttt{malloc}, each nodelet pre-allocates a local pool of edge blocks. A vertex can claim edge blocks from any pool, but it is desirable to string together edge blocks from the same pool to avoid thread migrations during edge list traversal. When the local pool is exhausted, the edge block allocator automatically moves to the pool on the next nodelet.

Kernel 1 of the Graph500 benchmark involves constructing a graph data structure from a list of edges. In our implementation the list of edges is loaded from disk into memory on nodelet 0. Currently I/O is limited on the prototype Emu Chick, and loading indirectly assists in evaluating the rest of the architecture. We sort the list by the low bits of the source vertex ID to group together edges that will be on the same nodelet, then spawn threads to scatter the list across all the nodelets. Once the list has been scattered, each nodelet spawns more threads locally to insert each edge into the graph, allocating edge blocks from the local pool.


\newcommand{\Neig}{\operatorname{Neig}}

\begin{table}
\centering 
\caption{Notations used in BFS.}
\begin{tabular}{r  l}
\textbf{Symbol} & \textbf{Description}                \\
\midrule
$V$             & Vertex set \\
$Q$             & Queue of vertices \\
$P$ & Parent array \\
$nP$             & New parent array \\
$\Neig(v)$     & Neighbor vertices of $v$\\
\midrule
\end{tabular}
\label{tab:TableOfNotationBFS}
\end{table}


\SetKwFor{Parfor}{for}{do in parallel}{endfor}

\begin{algorithm}
\begin{small}
  $P[v] \leftarrow -1$, for $\forall v\in V$\;
  $Q$.push(\textit{root})\;
  \While{$Q$ is not empty}{
    \Parfor{$s \in Q$}{
      \Parfor{$d \in \Neig(s)$}{\tcp*[l]{Thread migrates reading $P[d]$}
        \If{$P[d] = -1$}{
          \If{compare\_and\_swap($P[d]$, -1, $s$)}{$Q$.push($d$)}
        }
      }}
    $Q$.slide\_window()
  }
  \caption{BFS algorithm using migrating threads}
  \label{lst:bfs-migrating-threads}
\end{small}
\end{algorithm}

Our initial implementation of BFS (Algorithm~\ref{lst:bfs-migrating-threads}) was a direct port of the STINGER code. Each vertex iterates through each of its neighbors and tries to set itself as the parent of that vertex using an atomic compare-and-swap operation. If the operation is successful, the neighbor vertex is added to the queue to be explored along with the next frontier.

On Emu, the parent array is striped across nodelets in the same way as the vertex array. Each nodelet contains a local queue so that threads can push vertices into the queue without migrating. At the beginning of each frontier, threads are spawned at each nodelet to explore the local queues.  Thread migrations do occur whenever a thread attempts to claim a vertex that is located on a remote nodelet. In the common case, a thread reads an edge, migrates to the nodelet that owns the destination vertex, executes a compare-and-swap on the parent array, pushes into the local queue, and then migrates back to read the next edge. If the destination vertex happens to be local, no migration will occur when processing that edge.



\begin{algorithm}
\begin{small}
  $P[v] \leftarrow -1$, for $\forall v\in V$\;
  $nP[v] \leftarrow -1$, for $\forall v\in V$\;
  $Q$.push($root$)\;
  \While{$Q$ is not empty}{
    \Parfor{$s \in Q$}{
      \Parfor{$d \in \Neig(s)$}{
        \tcp*[l]{Thread issues remote write to $nP[d]$}
        $nP[d] \leftarrow s$}}
    cilk\_sync\;
    \Parfor{$v \in V$}{
      \If{$P[v] = -1$}{
        \If{$nP[v] \not= -1$}{
          $P[v] \leftarrow nP[v]$\;
          $Q$.push($v$)\;
          }}}
    $Q$.slide\_window()
  }

  \caption{BFS algorithm using remote writes}
  \label{lst:bfs-remote-writes}
\end{small}
\end{algorithm}

An alternative BFS implementation (Algorithm~\ref{lst:bfs-remote-writes}) exploits the capability of the Emu system to efficiently perform remote writes. A copy of the parent array ({\em nP}) holds intermediate state during each frontier. Rather than migrating to the nodelet that contains the destination vertex, we perform a remote write on the {\em nP} array. The remote write packet can travel through the network and complete asynchronously while the thread that created it continues to traverse the edge list. Remote writes attempting to claim the same vertex are serialized in the memory front end of the remote nodelet. Rather than attempting to synchronize these writes, we simply allow later writes to overwrite earlier ones. After all the remote writes have completed, we scan through the {\em nP} array looking for vertices that did not have a parent at the beginning of this frontier ($P[v] = -1$) 
but were assigned a parent in this iteration ($nP[v] \not= -1$). 
When such a vertex is found, it is added to the local queue, and the new parent value $nP[v]$ is copied into the parent array at $P[v]$.  This is similar to direction-optimizing BFS \cite{Beamer:2012:DBS:2388996.2389013} and may be able to adopt its early termination optimizations.
\newline

%

\subsection{{\sc\bfseries gsaNA}: Parallel Similarity Computation}
\label{ssec:gsana}
Integrating data from heterogeneous sources is often modeled as merging
graphs. Given two or more compatible, but not necessarily isomorphic graphs, the
first step is to identify a {\em graph alignment}, where a potentially
partial mapping of the vertices of the two graphs is computed. 
In this work, we investigate the parallelization of {\sc gsaNA}~\cite{yasar2018iterative}, which is
a recent graph aligner that uses the global structure of the graphs to
significantly reduce the problem space and align large graphs with a minimal
loss of information. The proposed
techniques are highly flexible, and they can be used to achieve higher recall
while being order of magnitudes faster than the current state of the art~\cite{yasar2018iterative}.

Briefly, {\sc gsaNA} first reduces the problem space, then runs pairwise
similarity computation between two graphs. Although the problem
space can be reduced significantly, the pairwise similarity computation step remains to be
the most expensive part (more than 90\% of the total execution time).
While {\sc gsaNA} has an embarrassingly parallelizable
nature for similarity computations, its parallelization is not
straightforward. This is because of the fact that {\sc gsaNA}'s similarity function is composed
of multiple components, with some only depending on graph structure and others depending 
also on the additional metadata (types and attributes). All of these components
compare vertices from two graphs and/or their neighborhood. Hence,
the similarity computation step has a highly irregular data access pattern. To reduce this
irregularity, we store the metadata of a vertex's neighborhood in sorted arrays.
While arranging metadata helps to decrease irregularity, data
access remains a problem because of the skewed nature of real-world graphs. Similarity 
computations require accessing different portions of the graph simultaneously.
In~\cite{yasar2018sina} authors provide parallelization strategies for different
stages of {\sc gsaNA}. However, because of the differences in the architecture and the
parallelization framework, the earlier techniques cannot be applied to  EMU Chick in a straightforward 
manner. Hence, in this work, 
we investigate two parallelization strategies for similarity computations and also two 
graph layout strategies on Emu Chick.

\begin{figure}
    \centering
    \begin{subfigure}[t]{0.3\linewidth}
        \centering
        \includegraphics[width=.9\linewidth]{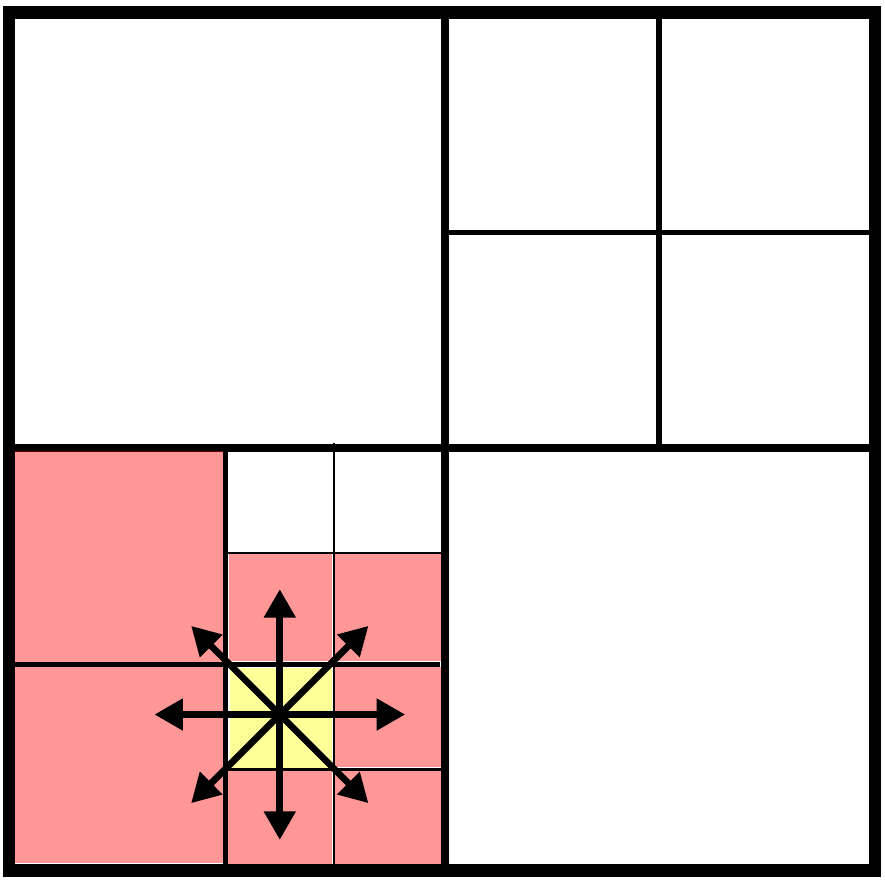}
        \caption{All}\label{fig:all}
    \end{subfigure}\hfill
    \begin{subfigure}[t]{0.3\linewidth}
        \centering
        \includegraphics[width=.9\linewidth]{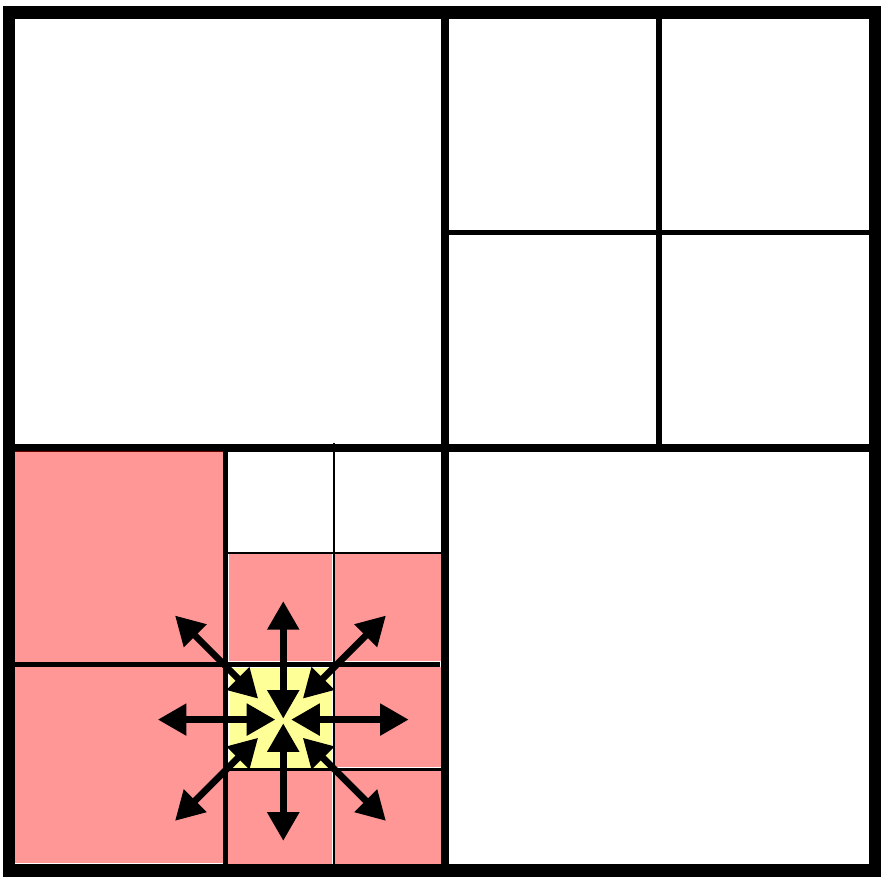}
        \caption{Pair}\label{fig:pair}
    \end{subfigure}\hfill
    \begin{subfigure}[t]{0.3\linewidth}
        \centering
        \includegraphics[width=.9\linewidth]{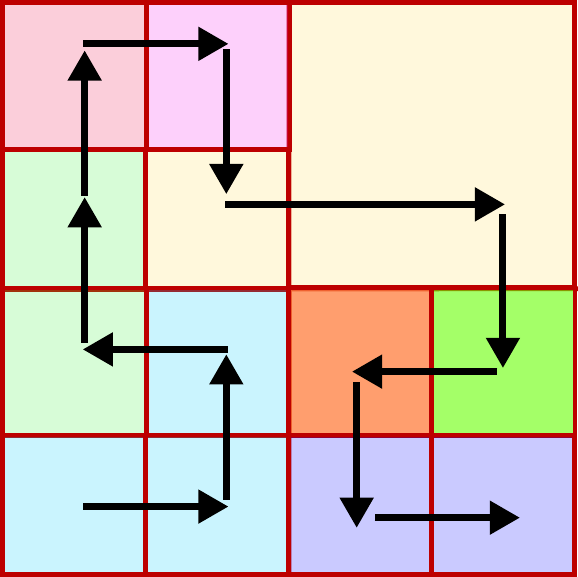}
        \caption{Layout}\label{fig:hcb}
    \end{subfigure}
    \caption{{\sc gsaNA}: {\small{Task definition \& bucket and vertex partition among the nodelets respecting the Hilbert-curve order.}}}
\end{figure}

\begin{table}[tbp]
\centering 
\caption{Notations used in {\sc gsaNA}.}
\begin{tabular}{r  l}
\textbf{Symbol} & \textbf{Description}                \\
\midrule
$V_1,V_2$             & Vertex sets \\
$E_1,E_2$             & Edge sets \\
$QT_1,QT_2$             & Quad-trees of the graphs \\
$QT_i.Neig(B)$     & Neighboring buckets of $B$ in $QT_i$\\
$\sigma(u,v)$   & Similarity score for $u\in V_1$ and $v \in V_2$ \\
$N(u)$        & Adjacency list of $u \in V_i$ \\
$A(u)$        & Vertex attribute of $u \in V_i$ \\
$RW(f(\cdot))$ & Number of required memory Reads \& Writes to  execute given function, $f(\cdot)$\\
\bottomrule
\end{tabular}
\label{tab:TableOfNotation}
\end{table}

{\sc gsaNA} places vertices into a $2D$ plane using a graph's global structure
information. The intuition is that similar vertices should also have similar
structural properties, and they should be placed closely on the $2D$ plane. When
all vertices are placed, {\sc gsaNA} partitions space into buckets in a quad-tree like fashion.
Then, a task for similarity computation becomes the pairwise comparison of the vertices in a
bucket with vertices in the neighboring buckets. For example, in Figures~\ref{fig:all}-\ref{fig:pair} the vertices in the yellow colored bucket are compared with vertices in the yellow and red colored buckets. 
We investigate two parallel similarity computation schemes and two vertex layout schemes.

\begin{algorithm}[ht]
\begin{small}
  $P[v] \leftarrow \emptyset$, for $\forall v\in V_2$\;

  \For{{\bf each} non-empty $B \in QT_2$}{
    {\tt cilk\_spawn} {\sc compSim}($B, QT_1.Neig(B), P, \sigma$)\;
  }
  {\tt cilk\_sync}\;
  \Return $P$
  \caption{{\sc parallelSim($QT_1,QT_2, k, \sigma$)}}
  \label{alg:par}
\end{small}
\end{algorithm}

\subsubsection{Similarity computation schemes.} In the {\em All Comparison} scheme, Alg.~\ref{alg:par} 
first spawns a thread for each non-empty bucket of $B \in QT_2$ where {\sc compSim} is instantiated with  {\sc compSimAll} shown in
Alg.~\ref{alg:csim}. This function computes the similarity
scores for each vertex $v \in B$ with vertex $u \in B'$ where
$B' \in QT_1.Neig(B)$. Afterwards, the top $k$ similar vertices are identified
and stored in $P[v]$. This approach has two main disadvantages. First, the number of
parallel tasks is limited by the number of buckets. Second, since space is
quad-tree like partitioned, this scheme may lead to load imbalance.
This technique is illustrated in Figure~\ref{fig:all}.
\begin{algorithm}[ht]
\begin{small}
  \tcp*[l]{For each vertex keep a priority list with top $k$ elements.}

  \For{{\bf each} $v \in B$} {
    \For{{\bf each} $B' \in N_B$} {
     \For{{\bf each} $u \in B'$ } {
      $P[v].insert(u)$  \tcp*[r]{Only keeps top $k$}      
      }
    }
  }
  \Return $P$
  \caption{{\sc compSimAll($B, N_B, P, \sigma$)}}
  \label{alg:csim}
\end{small}
\end{algorithm}

In the {\em Pair Comparison} scheme,
Alg.~\ref{alg:par}  
first spawns a thread for each non-empty bucket of $B \in QT_2$ where {\sc compSim} is instantiated with  {\sc compSimPair} shown in
Alg.~\ref{alg:csim2}.
Then, for each $\langle B, B' \rangle$ pair where $B \in QT_2$ and
$B' \in QT_1.Neig(B)$, {\sc compSimPairAux} is spawned. Next we
compute pairwise similarity scores of vertices between these bucket pairs
and return intermediate similarity scores (see Alg.~\ref{alg:csim2}). Finally,
we merge these intermediate results in Alg.~\ref{alg:csim2}.
This scheme spawns much more threads than the previous one.
This technique is illustrated in Figure~\ref{fig:pair}.

In ALL comparison scheme, the number of threads is limited by the number of buckets.Therefore achievable scalability is limited. On the other hand, because of the coarse grain composition of the tasks this scheme may lead to high load imbalance. Sorting tasks based on their loads in a non-increasing order can be a possible optimization/heuristic for reducing imbalance.

The PAIR comparison scheme reduces the load imbalance by compromising with additional
synchronization cost that arises during the insertion in Alg.~\ref{alg:csim}.
Task list is randomly shuffled to decrease the possibility of concurrent update
requests to a vertex's queue.

Note that while {\em ALL} is akin to vertex-centric based partitioning, 
{\em PAIR} is akin to edge-based partitioning. The vertices and edges here refer to the task graph.

\subsubsection{Vertex layouts.} In the {\em Block partitioned} (BLK) layout, the vertices are
partitioned among the nodelets based on their {\em ID}s, independent from their placement in the 2D plane. 
The buckets are also partitioned among the nodelets independently. That is, each nodelet stores an equal number of vertices and buckets. A vertex's metadata
is also stored in the same nodelet of corresponding vertex.
With the two computational schemes, vertices in the same bucket may be in different nodelets, leading to many thread migrations.
In the {\em Hilbert-curve based} (HCB) layout (shown in Fig.~\ref{fig:hcb}), the vertices and buckets
are partitioned among nodelets based on their Hilbert orders. To achieve this, after
all vertices are inserted in the quad-tree, we sort buckets based on their Hilbert orders.
Then, we rename every vertex in a bucket according to bucket's rank (i.e., vertices
in the first bucket, $B$, have labels starting from 0 to $|B|-1$). In
this layout every vertex is placed in the same nodelet with its bucket.
As with {\em BLK}, a vertex's metadata is also stored in the same nodelet
of the corresponding vertex.
Here, all vertices in the same bucket are in the same nodelet, and hence there is in general less migration.
While {\em BLK} may lead to a better workload balance (equal number of similarity computations per nodelet), {\em HCB} may lead
to a workload imbalance, if two buckets with high number of neighbors are placed into the same nodelet.

\begin{algorithm}[ht]
\begin{small}
  $p_{B'} \leftarrow \emptyset$, for $\forall B'\in N_B$\;

  \For{{\bf each} $B' \in N_B$} {
    {\tt cilk\_spawn} {\sc compSimPairAux}($B, B', p_{B'}, \sigma$ )
  }
  {\tt cilk\_sync}\;
  $P \leftarrow$ {\sc merge($p_{B' \in N_B}$)}\;
  \Return $P$
  \caption{{\sc compSimPair($B, N_B, P, \sigma$)}}
  \label{alg:csim2}
\end{small}

\begin{small}
  \;
{\bf def }{\sc compSimPairAux($B, B', P, \sigma$)}:\;
  \tcp*[l]{For each vertex keep a priority list with top $k$ elements.}

  \For{{\bf each} $v \in B$} {
    \For{{\bf each} $u \in B'$} {
      $P[v].insert(u)$  \tcp*[r]{Only keeps top $k$}
    }
  }
  \Return $P$
\end{small}
\end{algorithm}


\section{Experimental Setup}
\label{sec:exp}

\subsection{Emu Chick Prototype}
\label{ssec:prototype}
The Emu Chick prototype is still in active development. The current hardware iteration uses an Arria 10 FPGA on each node card to implement the Gossamer cores, the migration engine, and the stationary cores. Several aspects of the system are scaled down in the current prototype with respect to the next-generation system which will use larger and faster FPGAs to implement computation and thread migration. The current Emu Chick prototype has the following features and limitations:

\begin{compactitem}
\item Our system has one Gossamer Core (GC) per nodelet with a concurrent max of 64 threadlets. The next-generation system will have four GC's per nodelet, supporting 256 threadlets per nodelet.
\item Our GC's are clocked at 175MHz rather than the planned 300MHz in the next-generation Emu system.
\item The Emu's DDR4 DRAM modules are clocked at 1600MHz rather than the full 2133MHz.  Each node has a peak theoretical bandwidth of 12.8 GB/s.
\item CPU comparisons are made on a four-socket, 2.2 GHz Xeon E7-4850 v3 (Haswell) machine with 2 TiB of DDR4 with memory clocked at 1333 MHz (although it is rated for 2133 MHz). Each socket has a peak theoretical bandwidth of 42.7 GB/s.
\item The current Emu software version provides support for C++ but does not yet include functionality to translate Cilk Plus features like \texttt{cilk\_for} or Cilk \texttt{reducers}. All benchmarks currently use \texttt{cilk\_spawn} directly, which also allows more control over spawning strategies.

\end{compactitem}


\subsection{Experiment Configurations}

All experiments are run using Emu's 18.09 compiler and simulator toolchain, and the Emu Chick system is running NCDIMM firmware version 2.5.1, system controller software version 3.1.0, and each stationary core is running the 2.2.3 version of software. We present results for several configurations of the Emu system:

\begin{compactitem}
\item Emu Chick single-node \textbf{(SN)}: one node; 8 nodelets 
\item Emu Chick multi-node \textbf{(MN)}: 8 nodes; 64 nodelets
\item Simulator results are \emph{excluded from this study} as previous work \cite{hein2018emuchar} has shown simulated scaling numbers for SpMV and STREAM on future Emu systems. We prioritize multi-node results on hardware.
\end{compactitem}
\vspace{-1mm}


Application inputs are selected from the following sources:
\begin{compactitem}
  \item The SpMV experiments use synthetic Laplacian matrices, and real-world inputs are selected from the SuiteSparse sparse matrix collection \cite{davis2011university}. Each Laplacian consists of a five-point stencil which is a pentadiagonal matrix.
  \item BFS uses RMAT graphs as specified by Graph500 \cite{graph500} and uniform random (Erd\"os-Renyi) graphs \cite{1559977}, scale 15 through 21, from the generator in the STINGER codebase\footnote{\url{https://github.com/stingergraph/stinger/commit/149d5b562cb8685036517741bd6a91d62cb89631}}.
  \item {\sc gsaNA} uses DBLP~\cite{ref:dblp-site} graphs from years 2015 and 2017 that have been created previously~\cite{yasar2018iterative}. Detailed description of these graphs is provided in Section~\ref{ssec:res_gsna}.
\end{compactitem}

\subsection{Choosing Performance Metrics}
\label{sec:choos-perf-metr}

The Emu Chick is essentially a memory to memory architecture, so we primarily present results in terms of memory bandwidth and effective bandwidth utilization.
But comparing a new and novel processor architecture (Emu) built on FPGAs to a well-established and optimized architecture built on ASICs (Haswell) is difficult.
Measuring bandwidth on the Haswell with the STREAM benchmark\cite{McCalpin1995} achieves much more of the theoretical peak memory bandwidth.
The Emu Chick, however, implements a full processor on an FPGA and cannot take advantage of deeply pipelined logic that gives boosts to pure-FPGA accelerators, thus cannot achieve much of the theoretical hardware peak.
If we compare bandwidths against the DRAM peaks, prototype novel architectures like the Chick almost never appear competitive.
Comparing against measured peak bandwidth may provide an overly optimistic view of the prototype hardware.

We have chosen to primarily consider percentage of measured peak bandwidth given an idealized problem model, but also report the raw bandwidth results.
For integer SpMV and BFS, the natural measures of IOPS (integer operations per second) and TEPS (traversed edges per second) are proportional to the idealized effective bandwidth.

Our more recent tests have shown that the Emu hardware can achieve up to 1.6 GB/s per node and 12.8 GB/s on 8 nodes for the STREAM benchmark, which is used as the measured peak memory bandwidth number. This increase in STREAM BW from previous work \cite{hein2018emuchar} is primarily due to clock rate increases and bug fixes to improve system stability.  Meanwhile, our four-socket, 2.2GHz Haswell with 1333 MHz memory achieves 100 GB/s, or 25 GB/s per NUMA domain.  So the Emu FPGA-emulated processors achieve 11.7\% of the theoretical peak, while the ASIC Haswell processors achieve 58.6\%.  Note that we run with NUMA interleaving enabled, so many accesses cross the slower QPI links.  This provides the best Haswell performance for our pointer chasing benchmark\cite{hein2018emuchar}.  Disabling NUMA page interleaving brings the Haswell STREAM performance to 160 GB/s, which is 94\% of the theoretical peak.



\section{Results}
\label{sec:results}

\subsection{SpMV - to replicate or not, that is the question}
\label{ssec:res_spmv}

We first look at the effects of replication on the Emu, that is whether replicating the vector $x$ in Fig.~\ref{fig:csr_spmv_layout} provides a substantial benefit when compared to striping $x$ across nodelets in the ``no replication'' case.

{\bf Effective Bandwidth} is the primary metric measured in our experiments. It is calculated as the minimum number of bytes needed to complete the computation. On cache-based architectures this is equivalent to the compulsory misses.  For SpMV it is approximated by,
$$
BW = \frac{\texttt{sizeof(}A\texttt{)} + \texttt{sizeof(}x\texttt{)} + \texttt{sizeof(}y\texttt{)}}{\mathrm{time}}
$$
The numerator is a trivial lower-bound on data moved, since it counts only one load of $A$ (which enjoys no reuse) and one load each of the two vectors, $x$ and $y$ (assuming maximal reuse). The motivation for ignoring multiple loads of $x$ or $y$ is that ideally on a cache-based architecture with a ``well-ordered'' matrix, the vectors are cached and the computation is bandwidth-limited by the time to read $A$.
 
\begin{figure}
  \centering
  \includegraphics[width=.7\linewidth]{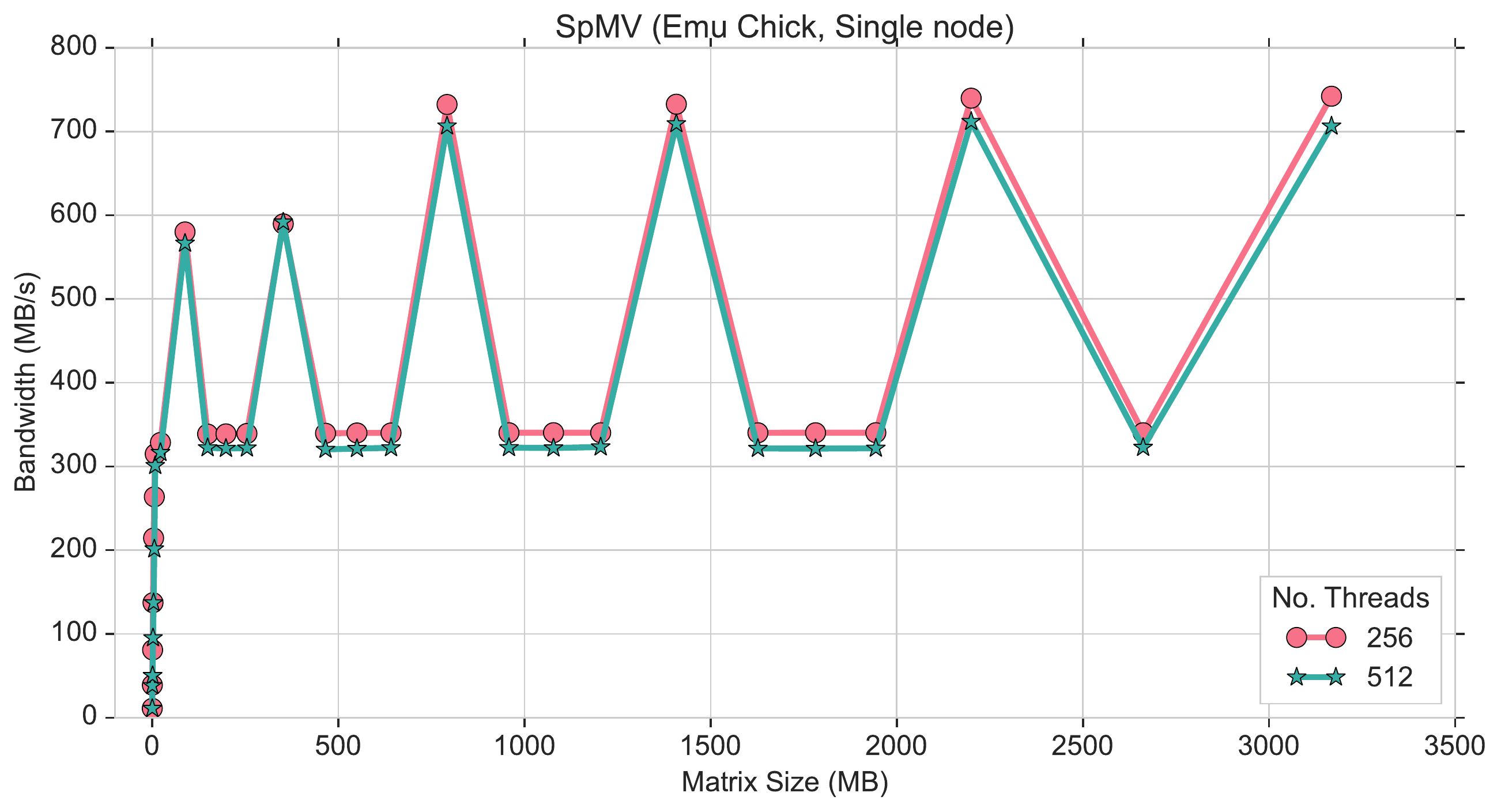}
  \caption{{\small{SpMV Laplacian Stencil Bandwidth, No Replicaton (8 nodelets).}}}\label{fig:spmv-nthreads}
\end{figure}

Figure~\ref{fig:spmv-nthreads} shows that the choice of grain size or work assigned to a thread can dramatically affect performance for the non-replicated test case. The unit of work here is the number of rows assigned to each thread. The default fixed grain size of 16, while competitive for smaller graphs, does not scale well to the entire node. For small grain sizes, too many threads are created with little work per thread, resulting in slowdown due to thread creation overhead. A dynamic grain size calculation is preferred to keep the maximum number of threads in flight, as can be seen with the peak bandwidth achieved with 256 and 512 threads for a single node. 

\begin{figure}
  \centering
  \includegraphics[width=.7\linewidth]{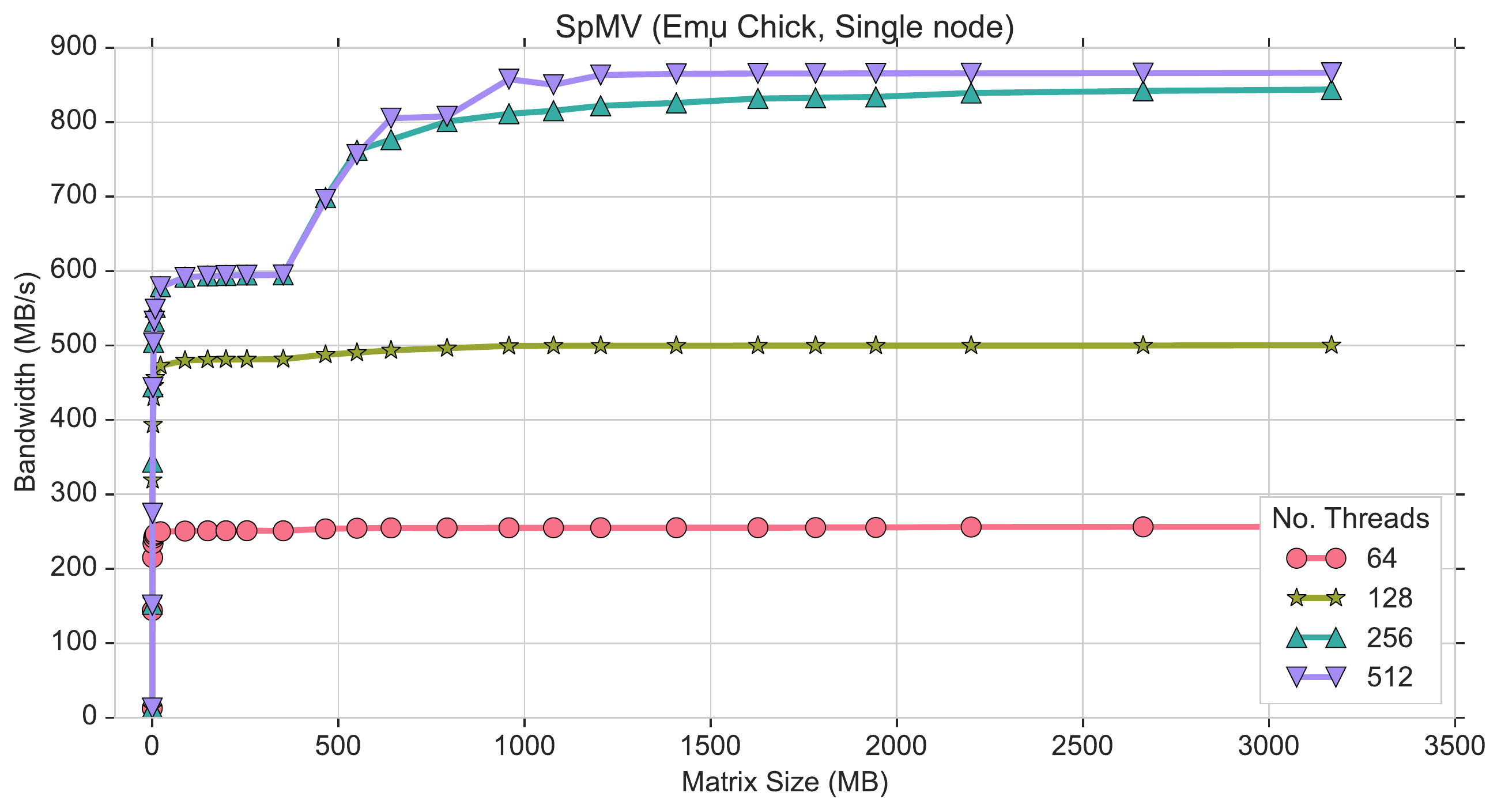}
  \caption{{\small{SpMV Laplacian Stencil Bandwidth, Replication (8 nodelets).}}}
  \label{fig:spmv-repl}
\end{figure}

Figure \ref{fig:spmv-repl} shows the effects of replication in SpMV.  Interestingly, for the largest matrix size both Figures \ref{fig:spmv-nthreads}~and~\ref{fig:spmv-repl} have similar bandwidths, which indicates good scaling for larger data sizes without replication at the potential cost of thread migration hotspots on certain nodelets. However, we note that using replication leads to much more regular scaling curves across different numbers of threads and grain sizes.

\begin{figure}
\centering
\includegraphics[width=.7\linewidth]{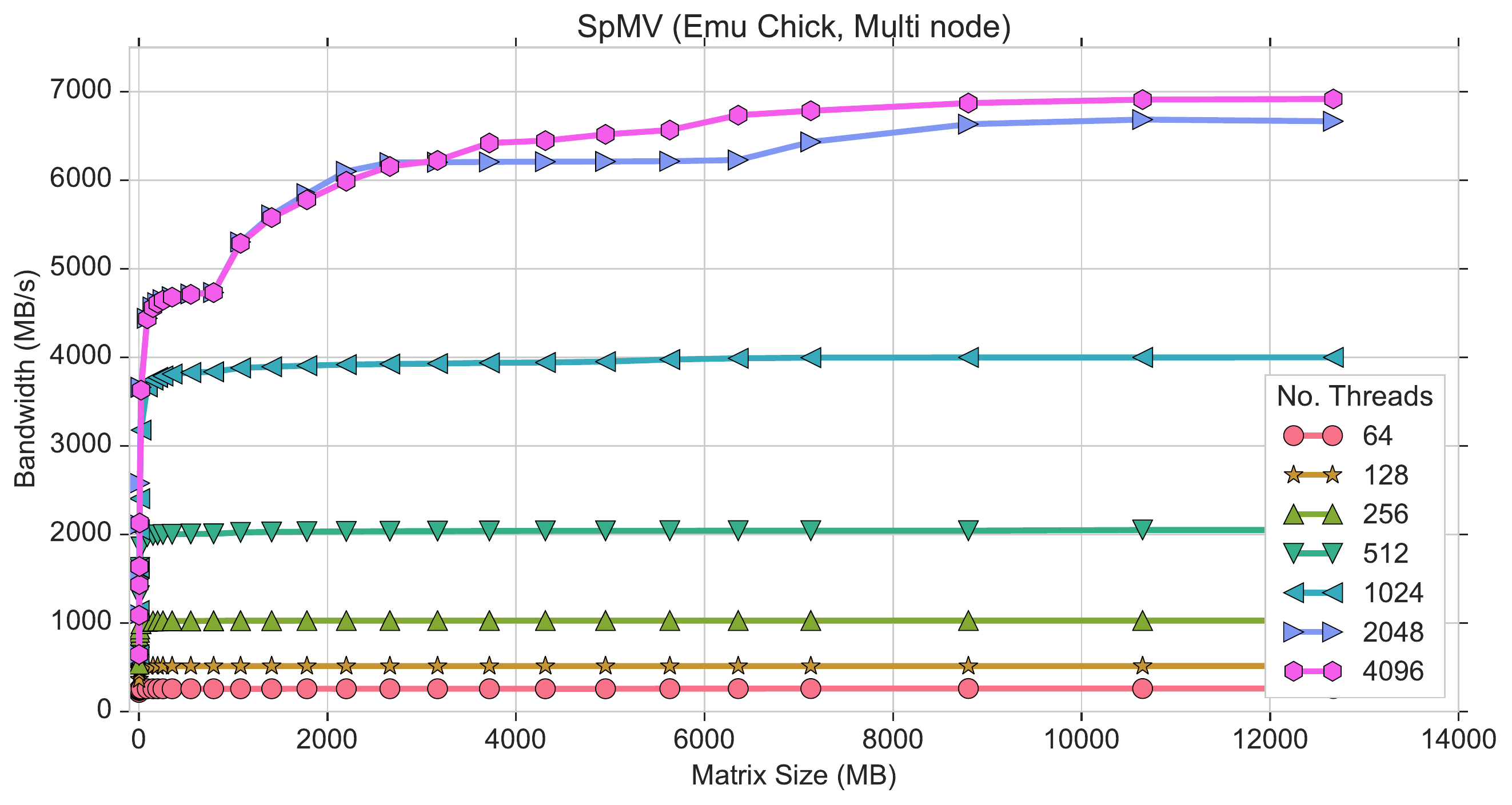}
\caption{{\small{SpMV Laplacian Replicated - multinode (64 nodelets).}}}
\label{fig:spmv-rep-multinode}
\end{figure}

Figure~\ref{fig:spmv-rep-multinode} shows scaling of multi-node (64 nodelets) using replication and different numbers of threads.
The best run of SpMV achieves 6.11 GB/s with 4096 threads, which is 50.8\% of our updated STREAM bandwidth number. However, it should also be noted from this figure that the best scalability for all sizes (including smaller inputs) is achieved using 2048 threads.

\begin{table}
\centering
\caption{SpMV multinode bandwidths (in MB/s) for real world graphs \cite{davis2011university} along with matrix dimension, number of non-zeros (NNZ), and the average and maximum row degrees. Run with 4K threads}
\label{table:spmv-real-multinode}
\begin{tabular}{lccccc}
\toprule
   Matrix &   Rows &    NNZ &  Avg Deg &  Max Deg &      BW \\
\midrule
  mc2depi &   526K &   2.1M &     3.99 &        4 & 3870.31 \\
 ecology1 &   1.0M &   5.0M &     5.00 &        5 & 4425.61 \\
 amazon03 &   401K &   3.2M &     7.99 &       10 & 4494.79 \\
 Delor295 &   296K &   2.4M &     8.12 &       11 & 4492.47 \\
 roadNet- &  1.39M &  3.84M &     2.76 &       12 & 3811.57 \\
 mac\_econ &   206K &  1.27M &     6.17 &       44 & 3735.54 \\
 cop20k\_A &   121K &  2.62M &    21.65 &       81 & 4520.05 \\
 watson\_2 &   352K &  1.85M &     5.25 &       93 & 3486.30 \\
   ca2010 &   710K &  3.49M &     4.91 &      141 & 4075.97 \\
 poisson3 &    86K &  2.37M &    27.74 &      145 & 4031.20 \\
   gyro\_k &    17K &  1.02M &    58.82 &      360 & 2446.36 \\
 vsp\_fina &   140K &   1.1M &     7.90 &      669 & 1335.59 \\
 Stanford &   282K &  2.31M &     8.20 &    38606 &  287.82 \\
     ins2 &   309K &  2.75M &     8.89 &   309412 &   43.91 \\
\bottomrule
\end{tabular}
\end{table}

Table~\ref{table:spmv-real-multinode} shows the multi-node (run with 2048 threads) bandwidth in MB/s for real-world graphs along with their average and maximum degree (non-zero per row) values. The rows are sorted by maximum degree and if we exclude the graphs with large maximum degree $(\geq 600)$ we see similar bandwidths. Most graphs showed bandwidths in excess of 600 MB/s and many were comparable to that of the synthetic Laplacians which are very well structured. This behavior is in contrast to a cache based system where we expect performance to increase with increasing degree. The Emu hardware demonstrates good performance independent of the structure of the graph, even ones with high-degree vertices.

For the high maximum degree graphs (\textit{Stanford}, \textit{ins2}) we attribute the poor performance to load imbalance. Some of the rows in these graphs have a very high number of non zeros. Since we only partition work at the row level, a single thread will need to process these large rows and this load imbalance results in slow running times. Current hardware limitations prevent exploring mixing parallelism across and within matrix rows \cite{7013032} leaving that level of performance benefit to future work.

\subsection{Graph500 - Migrating versus Remote Writes}
\label{ssec:graph500}

\begin{figure}[th]
  \centering
  \includegraphics[width=.7\linewidth]{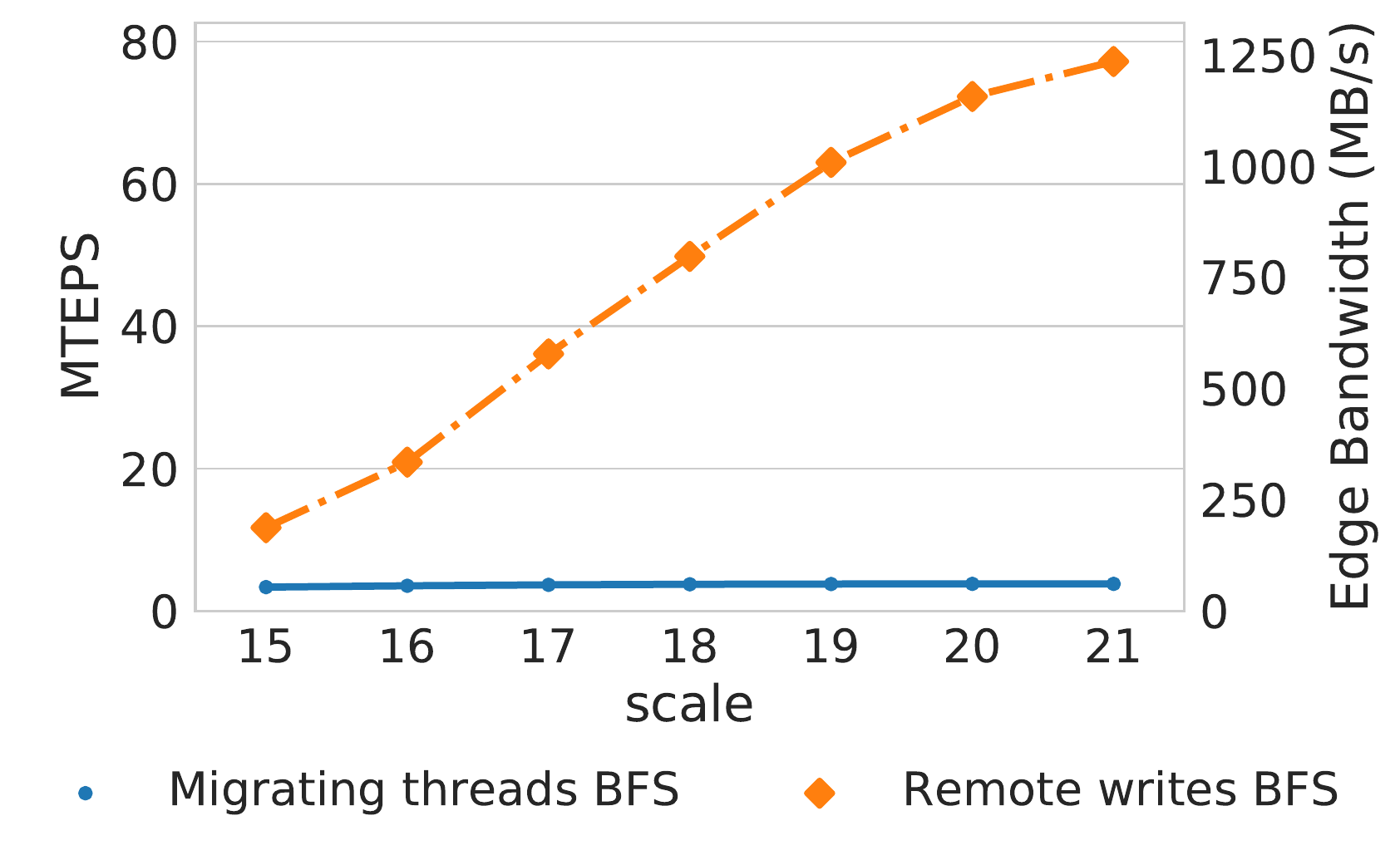}
  \caption{{\small{Comparison of remote writes versus migrating BFS on a multi-node Chick system for balanced (Erd\"os-R\'enyi) graphs. Marking members of the frontier with remote writes is more efficient than moving entire thread contexts back and forth between the edge list and the parent array. }}}
  \label{fig:emusim-bfs-mteps-compare-bfs-modes}
\end{figure}





Figure~\ref{fig:emusim-bfs-mteps-compare-bfs-modes} compares the migrating threads and remote write BFS implementations for a ``streaming'' or unordered BFS implementation. With the migrating threads algorithm, each thread will generally incur one migration per edge traversed, with a low amount of work between migrations. The threads are blocked while migrating, and do not make forward progress until they can resume execution on the remote nodelet. In contrast, the remote writes algorithm allows each thread to fire off many remote, non-blocking writes, which improves the throughput of the system due to the smaller size of remote write packets. 

The effective bandwidth for BFS on a graph with a given scale and an edge factor of 16 is as follows:
\begin{equation*}
  BW = \frac{16 \times 2^{\text{scale}} \times 2 \times 8}{\text{time}} = TEPS \times 2 \times 8.
\end{equation*}
This does not include bandwidth for flags or other state data structures and so is a lower bound as discussed in Section~\ref{ssec:res_spmv}.

\begin{figure}
  \centering
  \includegraphics[width=.7\linewidth]{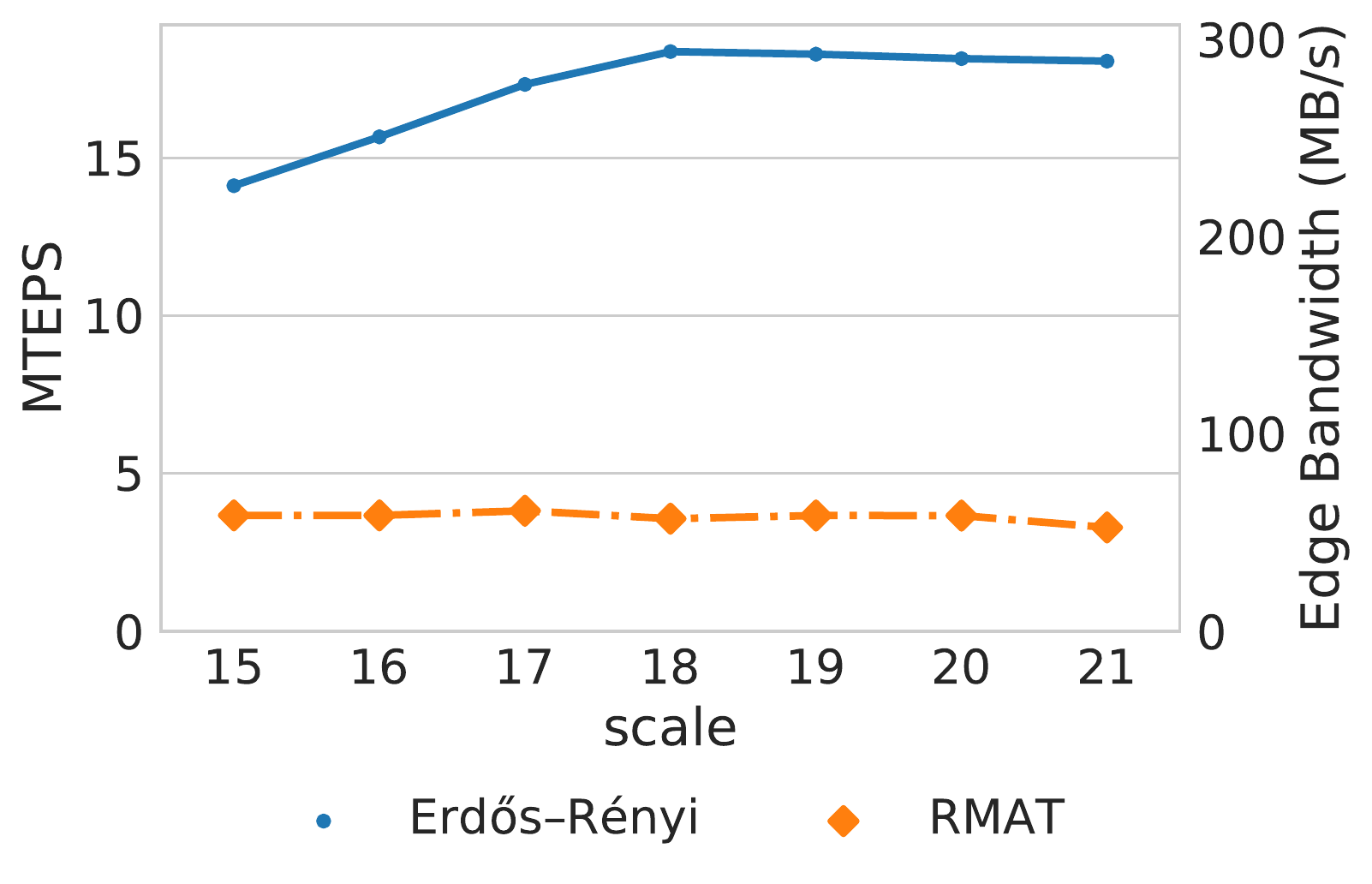}
  \caption{Compares the performance of BFS on a single-node system between balanced (Erd\"os-R\'enyi) graphs and unbalanced (RMAT) graphs running on a single node of the Emu Chick. Unbalanced graphs lead to an uneven work distribution and low performance.}
  \label{fig:emusim-bfs-mteps-compare-graph-types}
\end{figure}

Our initial graph engine implementation does not attempt to evenly partition the graph across the nodelets in the system. The neighbor list of each vertex is co-located with the vertex on a single nodelet. RMAT graphs specified by Graph500 have highly skewed degree distributions, leading to uneven work distribution on the Emu. Figure~\ref{fig:emusim-bfs-mteps-compare-graph-types} shows that BFS with balanced Erd\"os-R\'enyi graphs instead leads a higher performance of 18 MTEPS (288 MB/s) versus 4 MTEPS (64 MB/s) for the RMAT graph. We were unable to collect BFS results for RMAT graphs on the multi-node Emu system due to a hardware bug that currently causes the algorithm to deadlock. Future work will enhance the graph construction algorithm to create a better partition for power-law graphs.

\begin{figure}
  \centering
  \includegraphics[width=.7\linewidth]{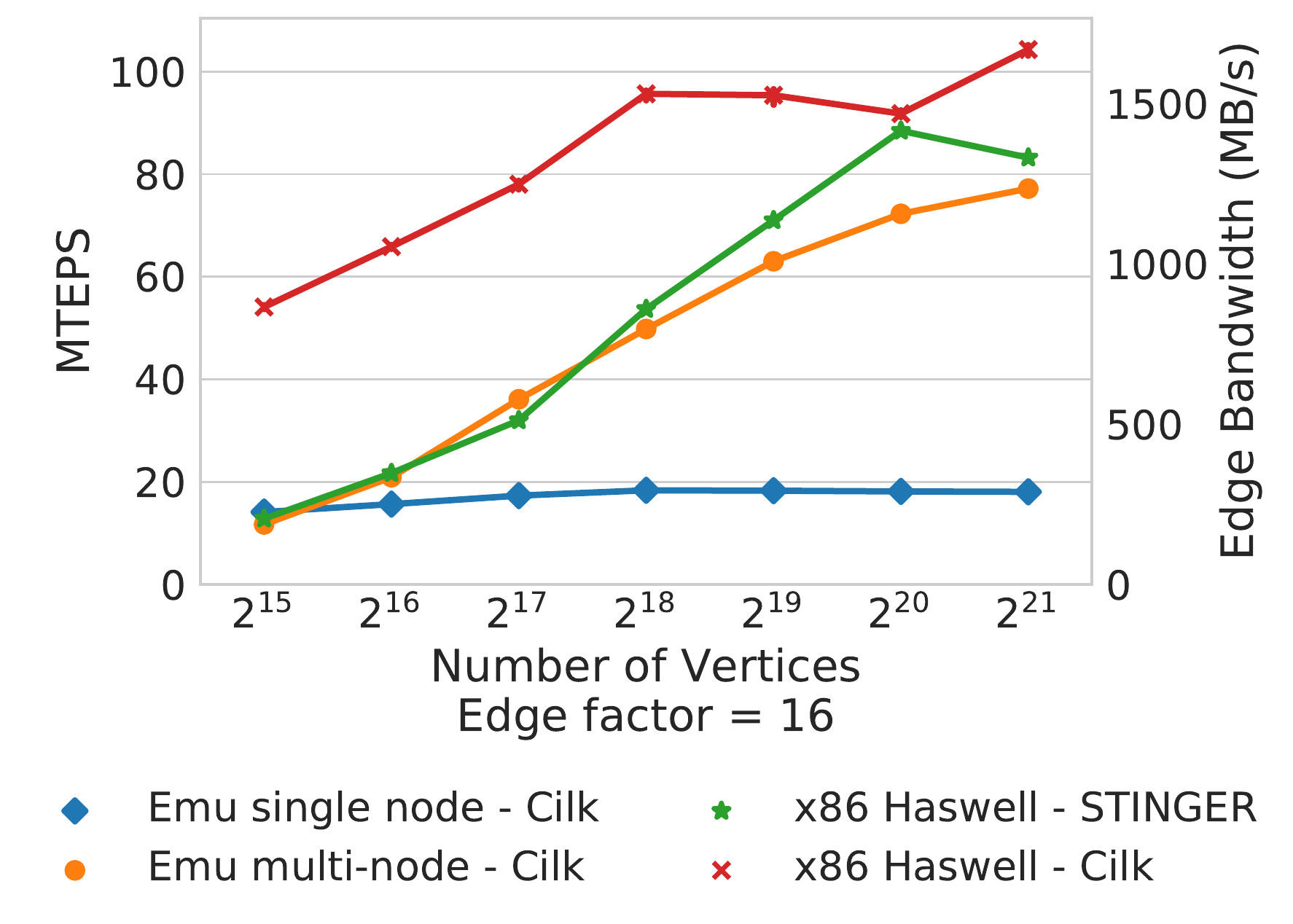}
  \caption{Comparison of BFS performance between the Emu Chick and the Haswell server described in Section~\ref{sec:exp}. Two implementations were tested on the Haswell System, one from STINGER and the other from MEATBEE}
  \label{fig:emu-bfs-vs-x86}
\end{figure}

Figure~\ref{fig:emu-bfs-vs-x86} plots results for four configurations of BFS running with balanced graphs: Emu single- and multi-node and two BFS results from the Haswell system. The performance of a single node of the Emu
Chick saturates at 18 MTEPS while the full multi-node configuration reaches 80 MTEPS on a scale 21 graph, with an equivalent bandwidth utilization of 1280 MB/s. On the Haswell platform, the MEATBEE (backported Emu Cilk) implementation reaches a peak of 105 MTEPS, outperforming the STINGER (naive Cilk) implementation of BFS at 88 MTEPS, likely due to the reduction of atomic operations as discussed in Section \ref{ssec:bfs_emu_alg}.

\subsection{ {\sc\bfseries gsaNA} Graph Alignment - Data Layout}
\label{ssec:res_gsna}

\begin{figure}
  \centering
  \includegraphics[width=.8\linewidth]{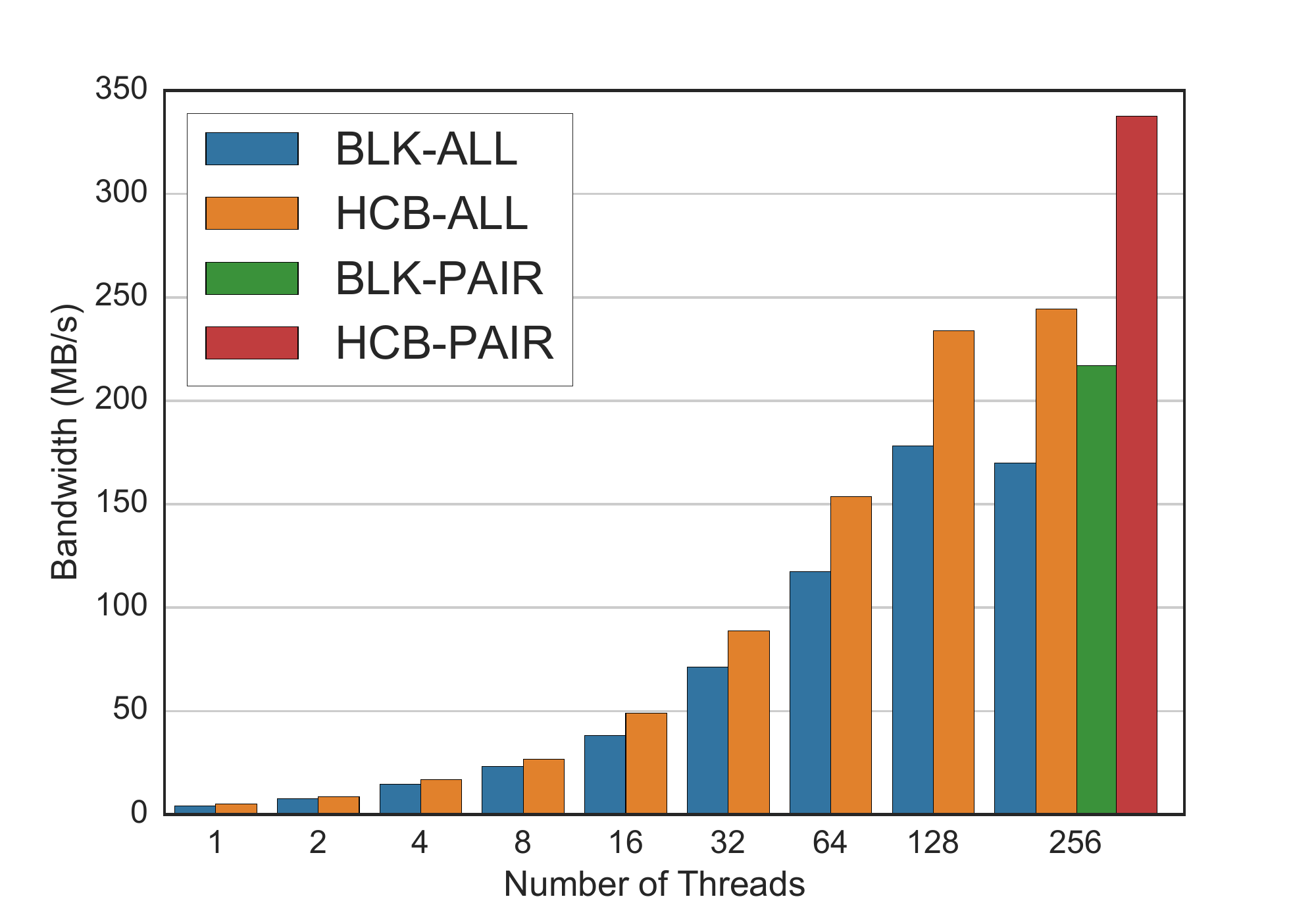}
  \caption{{\sc gsaNA}, {\small{Bandwidth vs. Threads for {\em ALL} (rightmost bars
   represent {\em PAIR} results), run on HW (8 nodelets).}}}
  \label{fig:gsana-phy}
\end{figure}

\begin{figure}
    \centering
	\includegraphics[width=.8\linewidth]{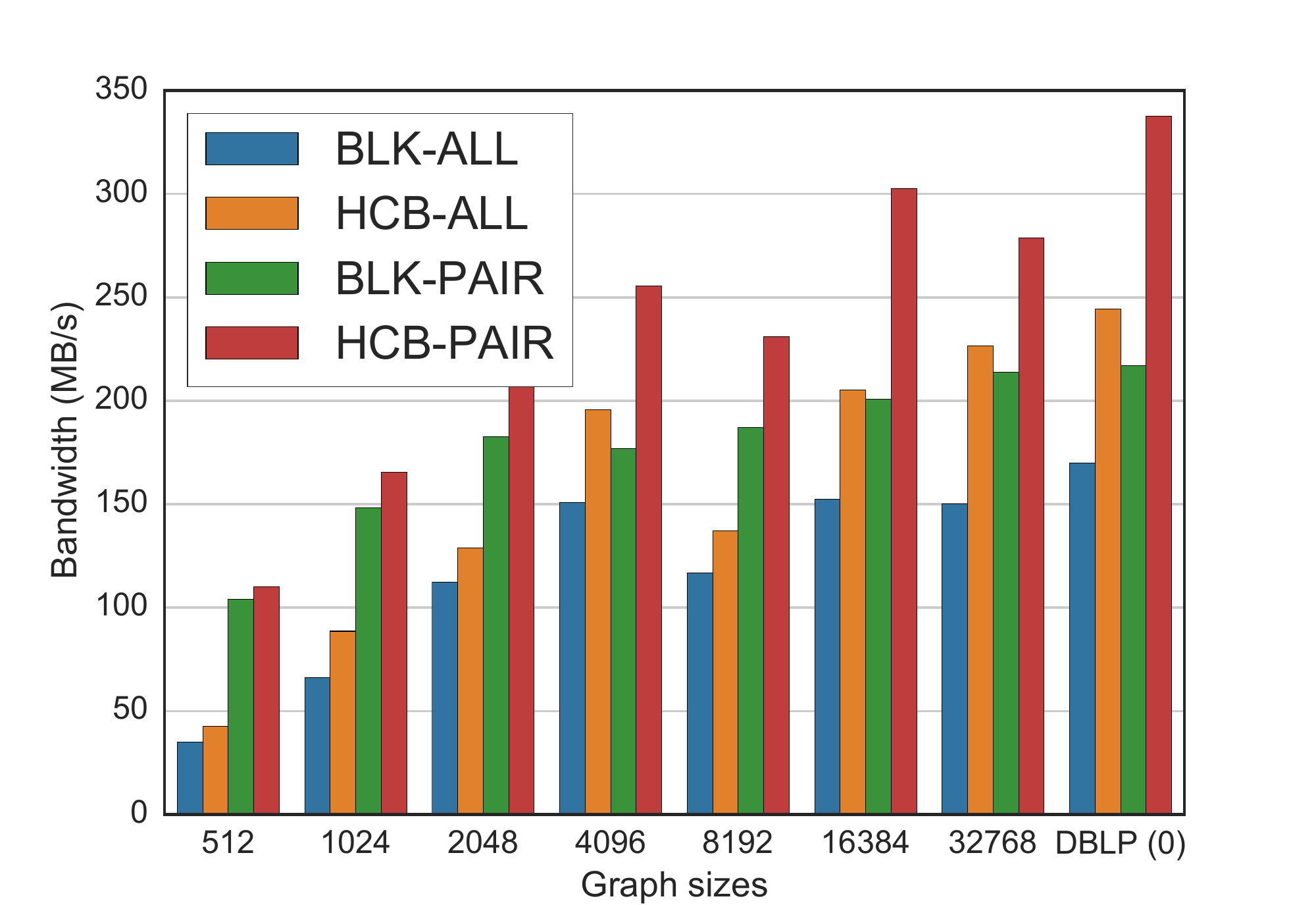}
    \caption{{\sc gsaNA}, {\small{Experiments on DBLP graphs on HW (8 nodelets).}}}
	\label{fig:gsana-phy-data}
\end{figure}

For our tests, we use DBLP~\cite{ref:dblp-site} graphs from years
$2015$ and $2017$ that have been created previously~\cite{yasar2018iterative}. This pair of graphs is called DBLP (0), and they have nearly $48K$, $59K$ vertices and $453K$, $656K$ edges
respectively. These graphs are used in the experiments shown in
Fig.~\ref{fig:gsana-phy}. For the experiments shown in
Fig.~\ref{fig:gsana-phy-data}, we filter some vertices
and their edges from the two graphs in DBLP (0), resulting in seven different graph pairs for alignment. The properties of these seven pairs are shown in
Table~\ref{table:gsana-dataset}.

\begin{table}
\centering
\caption{Generated graphs for alignment. $K=1024$;  $|T|$: number of tasks; $|B|$: bucket size.}
\label{table:gsana-dataset}
\begin{tabular}{l*{6}{c}r}
\toprule
 Graphs: & 512 & 1024 & 2048 & 4096 & 8192 & 16384 & 32768 \\
\midrule
$|V_1|,|V_2|$ & $0.5K$ & $K$  & $2K$ & $4K$ & $8K$ & $16K$ & $32K$ \\
$|E_1|$ 	& $1.3K$ & $4.4K$ & $14K$ & $35K$ & $88K$ & $186K$ & $385K$ \\
$|E_2|$		& $1K$ & $3K$ & $15K$ & $30K$ & $69K$ & $147K$ & $310K$ \\
\midrule

$|T|$ 	&$44$ &$85$ &$77$ &$163$ &$187$ &$267$ &$276$\\
$|B|$ 	&$32$ &$32$ &$64$ &$64$ &$128$ &$128$ &$256$\\
\bottomrule
\end{tabular}
\end{table}

We present similarity computation 
results for the Emu hardware on different sized graphs and execution schemes
which are defined/named by combining the
layout with the similarity computation. For instance,
{\em BLK-ALL} refers to the case where we use the block partitioned vertex layout and 
run {\em ALL} parallel similarity computation. {\bf Bandwidth} is measured
for {\sc gsaNA} by the formula:

\begin{align*}
BW &= \sum_{\forall B \in QT_1} \sum_{\substack{\forall B' \in \\QT_2.Neig(B)}} \frac{ \big(|B| + |B||B'| + \sum_{\forall u\in B} \sum_{\forall v \in B'} RW(\sigma(u,v))\big) \times \texttt{sizeof}(u) }{ \texttt{time}} 
\end{align*}

In a task, pairwise vertex similarities are computed between the vertices in a bucket $B\in QT_1$ and
the vertices in a bucket $B'\in QT_2.Neig(B)$.
Therefore in each task, every vertex $u \in B$ is read once and every vertex $v \in B'$ is read $|B|$ 
times. Additional read and write cost comes from the similarity function $\sigma(u,v)$ that 
is called for every vertex pair $u,v$ with 
$u\in B$ and $v \in  B'$. Hence, the total data movement can be gathered by aggregating the size of 
the bucket reads and the size of the number of reads and writes required by the similarity 
function. Bandwidth is the ratio between the total data 
movement over the execution time. 
We adopted the following similarity metrics from {\sc gsaNA}~\cite{yasar2018iterative}:
degree ($\Delta$), vertex type ($\tau$), adjacent vertex type ($\tau_V$), adjacent edge
type ($\tau_E$), and vertex attribute ($C_V$).
Since the similarity function consists of four different similarity 
metrics, we can define the required number of reads and writes of the similarity function as $RW(\sigma(u,v)) = RW(\tau(u,v)) + RW(\delta(u,v)) + RW(\tau_V(u,v)) + 
RW(\tau_E(u,v)) + RW(C_V(u,v))$. In this equation, the degree ($\Delta$) and the type ($\tau$) 
similarity functions 
require one memory read for each vertex and then one read and update for the similarity value.
Therefore, $RW(\tau(u,v)) = RW(\Delta(u,v)) = 4$. The adjacent vertex ($\tau_V$) and the edge ($\tau_E$) type 
similarity 
functions require reading all adjacent edges of the two vertices and one read and update for the 
similarity value. Therefore, $RW(\tau_V(u,v)) = RW(\tau_E(u,v)) = |N(u)| + |N(v)| + 2$. The vertex 
attribute similarity function ($C_V$) requires reading attributes of the two vertices and one 
read and update for the similarity value. Therefore, $RW(C_V(u,v)) = |A(u)| + |A(v)| + 2$.

The last three similarity metrics from {\sc gsaNA}~\cite{yasar2018iterative}
require comparing the neighborhood of two vertices, which causes a significant number
of thread migrations if the two vertices appear in different
nodelets. Therefore, these metrics are
good candidates to test the capabilities of the current hardware.

Figure~\ref{fig:gsana-phy} displays the bandwidth results of the similarity 
computation schemes for increasing numbers of threads, in different
execution schemes. In these
experiments, we only present results of the {\em PAIR} computation scheme
with the largest number of threads. Since the {\em PAIR} scheme
does many unpredictable recursive spawns, controlling the number of threads 
for this scheme is very hard and not accurate. Therefore, for increasing 
number of threads,
we only consider {\em ALL} with {\em BLK} and
{\em HCB} vertex layouts. We observe that in the {\em BLK} layout,
our final speedup is $43\times$ using {\em ALL} and 
$52 \times$ using {\em PAIR}. In the {\em HCB} layout,
our final speedup is $49 \times$ using {\em ALL}
and $68 \times$ using {\em PAIR}. As can be seen in 
Fig.~\ref{fig:gsana-phy}, when we increase the number of threads 
from $128$ to $256$, the
bandwidth decreases by $4\%$ in {\em BLK-ALL} scheme, because the coarse
grained nature of {\em ALL} cannot give better workload balance and
thread migrations hurt the performance.

Figure~\ref{fig:gsana-phy-data} presents results for all graphs in 
different execution schemes. We observe that the {\em HCB} vertex layout 
improves the execution
time by $10$-to-$36\%$ in all datasets by decreasing the number of
thread migrations. As expected, this improvement
increases with the graph size. This improvement in a x86 architecture
is reported as $10\%$ in~\cite{yasar2018sina}.
Second, we see that the {\em PAIR} 
computation scheme enjoys improvements with both vertex layouts, because it 
has a finer grained task parallelism and hence better workload distribution.

\begin{figure}
    \centering
	\includegraphics[width=.8\linewidth]{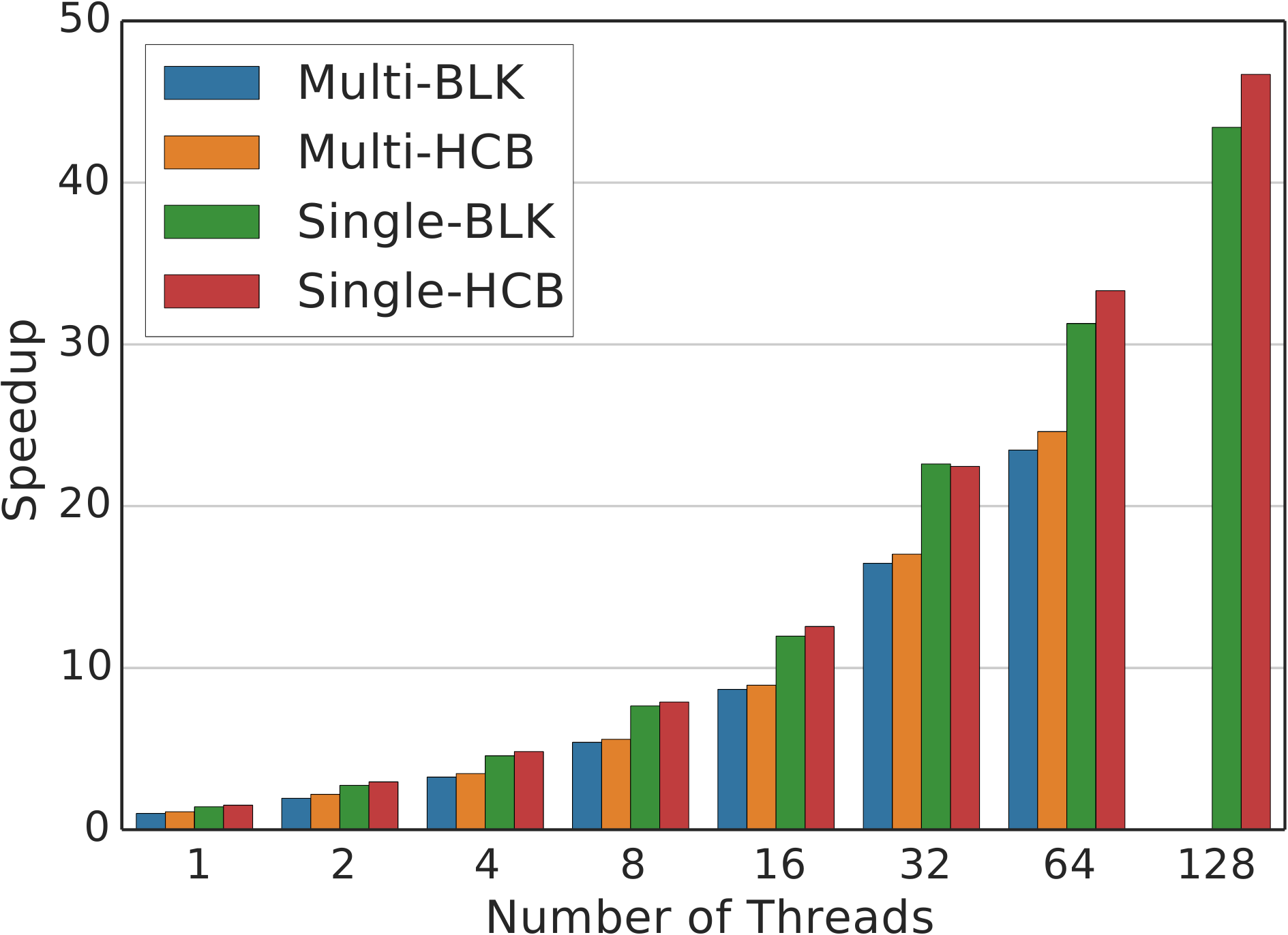}
    \caption{{\sc gsaNA}, {\small{strong scaling experiments on DBLP graph (2048 vertices) on HW (Multi-node and single-node).}}}
	\label{fig:gsanastrongscaling}
\end{figure}

Figure~\ref{fig:gsanastrongscaling} displays strong scaling results for BLK and HCB vertex
layouts with the {\em ALL} scheme on single-node and multi-node setups for the DBLP graph with $2048$ vertices.
Here, the strong scaling is given with respect to the single thread execution time of the BLK layout on the
multi-node setup. On the multi-node setup, hardware crashed for {\sc gsaNA} when $128$ threads were used.
We observe from this figure that multi-node setup is slower than the single node setup---multi-node execution times are about $25\%$ to
$30\%$ slower than the single-node execution times. This is so as the
inter-node migrations are much more expensive. The proposed layout and computational schemes help to improve
efficiency of the algorithms on both multi-node and single-node experiments. HCB layout
improves ALL layout about $12\%$ to $3\%$.

{\bf Final observations:} We observe that the finer granularity of tasks in 
{\em PAIR} and locality aware vertex layout with {\em HCB} give an 
important improvement in terms of the bandwidth (i.e., execution time). 
However, because of 
recursive spawns {\em PAIR} may cause too many unpredictable thread 
migrations
if the data layout is random. Additionally, although {\em HCB} helps to 
reduce the number of thread migrations significantly, this layout may 
create hotspots if it puts many neighboring buckets into the same 
nodelet. Our approach of balancing the number of edges per nodelet tries to alleviate these issues.


\subsection{Comparison to CPU-based Systems}
In addition to Figure \ref{fig:emu-bfs-vs-x86}, we present the following initial comparisons for our applications with runs using the same Cilk code on a Haswell CPU system. SpMV real-world graphs (Table \ref{table:spmv-real-multinode}) run in 5.33 ms to 1017.68 ms on the Emu versus 1.89 ms to 15.36 ms on the Haswell box, but the Emu exhibits memory bandwidth utilization of 0.3\%--38.53\% (where \textit{ins2} is slowest and \textit{amazon03} is fastest) versus 3.5--24\% on the CPU system. Emu results are skewed by the three most ``unbalanced'' graphs which run much faster on the CPU system. {\sc{gsaNA}} takes 158 seconds to run the DBLP(0) graph on the Emu hardware with $8$ nodelets and it takes less than $2$ seconds on a Haswell processor with $112$ threads. However, this is not a fair comparison as the presented DBLP graphs ($\sim$42 MB) fit in an L3 cache.

\section{Related Work}
\label{sec:related-work}

Advances in memory and integration technologies provide opportunities for profitably moving computation closer to data~\cite{Siegl:2016:DCF:2989081.2989087}.
Some proposed architectures return to the older processor-in-memory (PIM) and ``intelligent RAM''~\cite{592312} ideas.
Simulations of architectures focusing on near-data processing~\cite{7429299} including in-memory~\cite{8013497} and near-memory~\cite{7056040} show great promise for increasing performance while also drastically reducing energy usage.

Other hardware architectures have tackled massive-scale data analysis to differing degrees of success.
The Tera MTA / Cray XMT\cite{5161108,GraphCT-Wiley-Chap} provide high bandwidth utilization by tolerating long memory latencies in applications that can produce enough threads to source enough memory operations.
In the XMT all memory interactions are remote incurring the full network latency on each access.
The Chick instead moves threads to memory on reads, assuming there will be a cluster of reads for nearby data.  The Chick processor needs to tolerate less latency and need not keep as many threads in flight.
Also, unlike the XMT, the Chick runs the operating system on the stationary processors, currently PowerPC, so the Chick processors need not deal with I/O interrupts and highly sequential OS code.
Similarly to the XMT, programming the Chick requires language and library extensions.
Future work with performance portability frameworks like Kokkos\cite{CarterEdwards20143202} will explore how much must be exposed to programmers.
Another approach is to push memory-centric aspects to an accelerator like Sparc M7's data analytics accelerator\cite{7091786} for database operations or Graphicionado\cite{7783759} for graph analysis.

Moving computation to data via software has a successful history in supercomputing via {Charm++}\cite{7013040}, which manages dynamic load balancing on distributed memory systems by migrating the computational objects. Similarly, data analysis systems like Hadoop moved computation to data when the network was the primary data bottleneck~\cite{Ananthanarayanan:2011:DDC:1991596.1991613}. The Emu Chick also is strongly related to other PGAS approaches and is a continuation of the mNUMA architecture~\cite{Vance:2010:IME:2020373.2020379}. Other approaches to hardware PGAS acceleration include advanced RDMA networks with embedded address translation and atomic operations \cite{6468518,5739087,7312665,Dungworth2011}. The Emu architecture supports remote memory operations to varying degrees and side-steps many other issues through thread migration.
Remote operations pin a thread so that the acknowledgment can be delivered.
How to trade
between migration and remote operations, as well as exposing that trade-off, is an open question.

\textbf{SpMV:} There has been a large body of work on SpMV including on emerging architectures \cite{williams2009optimization,bonachea2006efficient} but somewhat limited recent work that is directly related to PGAS systems. However, future work with SpMV on Emu will investigate new state-of-the-art formats and algorithms such as SparseX, which uses the Compressed Sparse eXtended (CSX) as an alternative data layout for storing matrices~\cite{Elafrou:2018:sparsex}.

\textbf{BFS:} The implemented version of BFS builds on the standard Graph500 code with optimizations for Cilk and Emu. The two-phase implementation used in this work has some similarities to direction-optimizing BFS \cite{6651058}, in that the remote "put" phase mirrors the bottom-up algorithm. Other notable current implementations include optimized, distributed versions~\cite{Ueno2017} and a recent PGAS version~\cite{Cong:2010:FPI:1884643.1884669}. The implementation provided in this paper contrasts with previous PGAS work due to asymmetric costs for remote get operations as discussed in Section \ref{sec:disc}. 
NUMA optimizations\cite{10.1007/978-3-319-07518-1_23} similarly are read-oriented but lack thread migration.

\textbf{Graph Alignment:} Graph alignment methods are traditionally~\cite{conte2004ijprai,elmsallati2016itcbb} 
classified into four basic categories: spectral methods
~\cite{singh2007pairwise,liao2009bioinf,patro2012bioinf,neyshabur2013bioinf,zhang2016kdd};
graph structure similarity methods
~\cite{kuchaiev2010topological,milenkovic2010optimal,vesna2012graal,dogning2015binf,aladag2013bioinf};
tree search or table search methods
~\cite{chindelevitch2013optimizing,saraph2014bioinf,liu2007multiobjective,kpodjedo2014using};
and integer linear programming methods
~\cite{klau2009bmc,kebir2011icprb,bayati2009icdm,koutra2013icdm}.
{\tt Final}~\cite{zhang2016kdd} is a recent work which targets labeled network alignment problem by extending the concept of
{\tt IsoRank}~\cite{singh2007pairwise} to use attribute information of the vertices and edges. 
All of these methods have
scalability issues. {\sc gsaNa}~\cite{yasar2018iterative,yasar2018sina}
leverages global graph structure and reduces the problem space
and exploits the semantic information to alleviate most of the
scalability issues. In addition to these sequential algorithms, we are
aware of two parallel approaches for global graph alignment. The first
one~\cite{kollias2012tkde}
decomposes the ranking calculations of {\tt IsoRank}'s
similarity matrix using the singular value decomposition.
The second one is a shared memory parallel algorithm~\cite{neyshabur2013netal}
that is based on the belief propagation (BP) solution for integer program
relaxation~\cite{bayati2009icdm}. It uses shared memory parallel matrix
operations for BP iterations and also implements an approximate weighted
bipartite matching algorithm. While these parallel algorithms show an
important improvement over the state of the art sequential algorithms,
the graphs used in the experiments are small in size and
there is a high structural similarity. To the best of our knowledge, the use of {\sc{gsaNA}} in 
~\cite{yasar2018sina} and in this paper presents the first method for parallel alignment of 
labeled graphs.

Other recent work has also looked to extend from low-level characterizations like those presented here by providing initial Emu-focused implementations of Breadth-First Search\cite{belviranli:2018:emuhpec}, 
Jaccard index computation 
\cite{krawezik:2018:jaccard_emu_hpec}, bitonic sort, \cite{velusamy:2018:sort_emu_hpec} and compiler optimizations like loop fusion, edge flipping, and remote updates to reduce migrations
\cite{chatarasi:2018:psc_emu_mchpc}.


\section{Emu Architectural Discussion} 
\label{sec:disc}


The Emu architecture inverts the traditional scheme of hauling data to and from a grid of processing elements. In this architecture, the data is static, and small logical units of computation move throughout the system, and the load balancing is closely related to data layout and distribution, since threads can only run on local processing elements.
Our work mapping irregular algorithms to the Emu architecture expose the following issues needed to achieve relatively high performance:
\begin{compactenum}
\item \textbf{Thread stack placement} and remote thread migrations back to a ``home'' nodelet that contains the thread stack. 
\item \label{disc:limit:balance} \textbf{Balancing workload} is difficult when using irregular data structures like graphs.
\item 
Following from \ref{disc:limit:balance}, input sizes are limited by the need to create \textbf{distributed data structures} from an initial chunk of data on the ``home'' node. 
\item The Emu is a \textbf{non-uniform PGAS} system with variable costs for remote ``put'' and ``get'' operations. 
\item The tension between \textbf{top-down task programming on bottom-up data allocation} has proven difficult to capture in current programming systems.
\end{compactenum}

\textbf{Thread Stack Placement:} A stack frame is allocated on a nodelet when a new thread is spawned. Threads carry their registers with them when they migrate, but stack accesses require a migration back to the originating nodelet. If a thread needs to access its stack while manipulating remote data, it will migrate back and forth (ping-pong). We can limit the usage of thread stacks and ping-pong migration by obeying the following rules when writing a function that is expected to migrate:
\begin{compactenum}
\item Maximize the use of inlined function calls. Normal function calls require a migration back to the home nodelet to save the register set.
\item Write lightweight worker functions using fewer than 16 registers to prevent stack spills. 
\item Don't pass arguments by reference to the worker function. Dereferencing a pointer to a variable inside the caller's stack frame forces a migration back to the home nodelet. Pointers to replicated data circumvent this migration.
\end{compactenum}

\textbf{Workload balance and distributed data structures:} One of the main challenges in obtaining good performance on the Emu Chick prototype is the initial placement of data and distribution to remote nodelets. While the current Emu hardware contains a hardware-supported credit system to control the overall amount of dynamic parallelism the choice of placement is still critical to avoid thread migration hotspots for SpMV and BFS. In the case of SpMV, replication reduces thread migration in each iteration, but replication is also not scalable to more complex, related algorithms like MTTKRP or SpGEMM. The implementations of graph alignment using gsaNA uses data placement techniques like HCB and PAIR-wise comparisons to group threads on the same nodelets with related data and limit thread migration, which dramatically improves their performance.

\textbf{Non-uniform PGAS operations:} Emu's implementation of PGAS utilizes ``put''-style remote operations (add, min, max, etc.) and ``get'' operations where a thread is migrated to read a local piece of data. Thread migration is efficient when many get operations need to access the same nodelet-local memory channel. The performance difference observed between put and get operations is due to how these two operations interact differently with load balancing. A put can be done without changing the location of the thread, while a get means that multiple threads may have to share significant resources on the same nodelet for a while. Additionally, a stream of gets with spatial locality can be faster than multiple put operations. This non-uniformity means that kernels that need to access finely grained data in random order should be implemented as put operations wherever possible while get operations should only be used when larger chunks of data are read together.  A major outstanding question is how this scheme compares with explicitly remote references plus task migrations via remote calls as in UPC++\cite{upcxxv8}.  The trade-off between hardware simplicity and software flexibility is difficult to measure without implementations of both.  Tractable abstract models miss implementation details like switch fabric traffic contention or task-switching cache overhead.

\textbf{Top-down task programming on bottom-up data allocation:}  The Cilk-based fork/join model emphasizes a top-down approach to maximize parallelism without regard to data or thread location.  Memory allocation on the Emu system, however, follows the bottom-up approach of UPC\cite{osti_1134233} or SHMEM\cite{Chapman:2010:IOS:2020373.2020375}.  The Cilk model allows quickly writing highly parallel codes, but achieving high performance (bandwidth utilization) requires controlling thread locations.  We do not yet have a good way to relieve these tensions.  Languages like Chapel\cite{Chamberlain:2007:PPC:1286120.1286123} and X10\cite{Charles:2005:XOA:1103845.1094852} provide a high-level view of data distribution but lack implicit thread migration.  The \textsc{gaANA} results on the highly dynamic variant in Algorithm~\ref{alg:csim2} demonstrate how migrations on locality-emphasizing data distribution can achieve relatively high performance.  To our knowledge there is little work on programming systems that incorporate \emph{implicit and light-weight} thread migration, but Charm++\cite{7013040} and Legion\cite{Bauer:2012:LEL:2388996.2389086} provide experience in programming heavier-weight task migration and locality in different environments.

Note that the Emu compiler is rapidly evolving to include intra-node \texttt{cilk\_for} and Cilk+ reducers.  Experimental support became available at the time of writing and still is being evaluated.  Balancing remote memory operations and thread migrations in reducer and parallel scan implementations for the Emu architecture is ongoing work.





\section{Conclusion} \label{sec:concl}
In this study, we focus on optimizing several irregular algorithms using programming strategies for the Emu system including replication, remote writes, and data layout and placement. We argue that these three types of programming optimizations are key for achieving good workload balance on the Emu system and that they may even be useful to optimize Cilk-oriented codes for x86 systems (as with BFS). 

By analogy, back-porting GPU-centric optimizations to processors often provides improved performance.  That is, in the same way that GPU architecture and programming encourages (or ``forces'') programmers to parallelize and vectorize explicitly, the Emu design requires upfront decisions about data placement and one-sided communication that can lead to more scalable code. Future work would aim to evaluate whether these programming strategies can be generalized in this fashion.

By adopting a "put-only" strategy, our BFS implementation achieves 80 MTEPS on balanced graphs. Our SpMV implementation makes use of replicated arrays to reach 50\% of measured STREAM bandwidth while processing sparse data. We present two parallelization schemes and two vertex layouts for parallel similarity computation with the {\sc gsaNA} graph aligner, achieving strong scaling up to $68 \times$ on the Emu system. Using the {\em HCB} vertex layout further improves the execution time by up to $36\%$.


\begin{acks}
This work partially was supported by NSF Grant ACI-1339745 (XScala), an IARPA contract, and the Defense Advanced Research Projects Agency (DARPA) under agreement \#HR0011-13-2-0001. Any opinions, findings, conclusions, or recommendations in this paper are solely those of the authors and do not necessarily reflect the position or the policy of the sponsors. Thanks also to the Emu Technology team for support and debugging assistance with the Emu Chick prototype.
\end{acks}


\bibliographystyle{ACM-Reference-Format}
\bibliography{bib/refs} 



\end{document}